# Automatic Metadata Extraction for Text-to-SQL

Vladislav Shkapenyuk
AT&T CDO
vs9593@att.com

Divesh Srivastava
AT&T CDO
ds8961@att.com

Theodore Johnson
AT&T CDO
tj1857@att.com

Parisa Ghane
AT&T CDO
pg4126@att.com

## ABSTRACT

Large Language Models (LLMs) have recently become sophisticated enough to automate many tasks ranging from pattern finding to writing assistance to code generation. In this paper, we examine text-to-SQL generation. We have observed from decades of experience that the most difficult part of query development lies in understanding the database contents. These experiences inform the direction of our research.

Text-to-SQL benchmarks such as SPIDER and Bird contain extensive metadata that is generally not available in practice. Human-generated metadata requires the use of expensive Subject Matter Experts (SMEs), who are often not fully aware of many aspects of their databases. In this paper, we explore techniques for automatic metadata extraction to enable text-to-SQL generation.

We explore the use of two standard and one newer metadata extraction techniques: profiling, query log analysis, and SQL-to-text generation using an LLM. We use BIRD benchmark [JHQY+23] to evaluate the effectiveness of these techniques. BIRD does not provide query logs on their test database, so we prepared a submission that uses profiling alone, and does not use any specially tuned model (we used GPT-4o). From Sept 1 to Sept 23, 2024, and Nov 11 through Nov 23, 2024 we achieved the highest score both with and without using the "oracle" information provided with the question set. We regained the number 1 spot on Mar 11, 2025, and are still at #1 at the time of the writing (May, 2025).

## 1 INTRODUCTION

Recent developments in Large Language Models (LLMs) have spurred extensive research into automating many activities, including code generation. In this paper, we address text-to-SQL generation [QHWY+22].

Our research was informed by our experiences with large-scale industrial databases [DJMS02][GJ14]. We have consistently found that the most difficult part of writing correct SQL queries is understanding what is in the database to begin with. After that, writing queries is relatively straightforward. Examples of difficulties include:

- **Lack of documentation:** In a large number of cases, the database lacks documentation – no field or table descriptions. Primary key – foreign key dependencies are not labeled. The new database user has only cryptic field and table names for guidance.
- **Incomplete or dated documentation:** Databases undergo continual schema and data change. Fields get added or dropped, and their contents can change their format and meaning. Often times, the documentation is not updated. In general, even the Subject Matter Experts (SMEs) are often not fully aware of the current contents of a database.
- **Unclear data formats**: To take a simple example, suppose that we know that the *owner* field is the name of the owner of the account. What is the format? Is it *last_name first_name*, *last_name, first_name*, or *first_name last_name*? Is there any capitalization? Punctuation? How are more than two names handled?
- **Multiple data formats**: A single field might have data in multiple formats. For example, the *owner* field might half its entries in format *first_name last_name* and half in format *last_name, first_name*.
- **Multiple fields with similar meanings:** A common example of this issue is a *date* field. Suppose that table *customer* has 4 date fields, *a_date, b_date, c_date, d_date* each of which represents a different date of an interaction (actual names can be almost as cryptic). How does one find all customers who signed up on or after Sept. 1, 2024? Different fields might be filled in based on the sign up process so the proper formula might be *Coalesce(a_date, c_date) >= date '01/09/2024'*.
- **Complex join paths:** There can be many complexities in developing a correct join expression – the join key might involve multiple fields, it might involve conditional values (e.g. because of multiple processes adding records to the table – *iff(R.vendor=5, R.id, R.serial_num)=S.serial_num*), or it might involve transformations on join keys. In one example, we needed to join two well-documented tables on the IMEI[1]. The join result was empty - until we realized that in one table the IMEI was 13 digits and in the other it was 14. The longer IMEI always had '1' as a prefix.

- **Complex formulae**: The complexities of an operations or business process can make seemingly simple calculations obscure. For example, the revenue of an interaction might be calculated as *iff(cds_code=1, revenue, zdc+total_rendered*10000)*.
- **Default values:** In many cases, seemingly required fields have nonsense values. For example, a telephone number field might have a very large number of entries such as 123-456-7890 or 111-111-1111. These default values should be excluded from results involving the telephone number. If the telephone number is a join key between two 1-million record tables and both fields have 123-456-7890 in one tenth of their records, then the query must materialize 10 billion useless records.

Several approaches exist for understanding complex databases. One well-known approach is *profiling* [AGN15], the systematic querying of the tables in a database to create reports on their properties. For example, the profiling might show that the T.IMEI field is always 13 digits while the S.IMEI field is always 14 digits and always starts with a 1. A natural conclusion is that T and S are joinable on '1'||T.IMEI=S.IMEI.

Another common approach is to examine queries that SMEs have written for clues to important fields, join paths, business logic, and so on. If a query log is available, this analysis can be automated to generate statistical reports which might indicate e.g. complex join paths [YPS09].

A third approach is newer, and is enabled by LLMs. If one has a query log and high-quality metadata, one can ask the LLM to translate the SQL into text. This technique allows the user to find related queries based on the textual similarity of the questions.

In this paper, we apply these techniques to LLM-based text-to-SQL generation. For evaluation, we use the BIRD benchmark [JHQY+23] to quantify their benefit. BIRD is a challenging benchmark, with often ambiguous and/or dirty schemas, data, and metadata. Our experiments are run on the *dev* database, with questions selected from *minidev* for some experiments.

In the BIRD benchmark, a submission is evaluated on an unknown *test* database – so no query log is available. We developed a submission which uses only profiling information. On Sept. 1, 2024 and again on Nov 11,2024, we achieved the highest scores both using and not using the *oracle*[2] information. Since oracle information is never present in practice, the test which does not use the oracle is more indicative of how a text-to-SQL technique works in practice. Without the oracle, our submission got a score 10.28 percentage points higher than the next best submission without the oracle, 67.41% vs. 57.13% (at the time of writing – Jan. 2025). On March 11, 2025 we submitted again using the oracle and achieved the #1 spot with a test score of 77.14. The top five scores using hints, at the time of writing, are: 77.14 (AT&T), 76.02 (Google), 75.63 (Contextual AI), 75.63 (Alibaba), 73.17 (IBM).

## 1.1 Contributions

In this paper, we investigate three schemes for automatic metadata generation for text-to-SQL applications. Two of these schemes are traditional: database profiling and query log analysis. In the context of database profiling, we show that an LLM can translate the profiling information (in the context of the table schema) into useful metadata about field meaning.

We use the BIRD benchmark to evaluate automatic metadata extraction. In the context of database profiling, we find that by using the LLM to summarize the profile metadata, we can gain significant insights into field contents.

Using profile-enhanced field metadata blows up the size of the schema provided in an LLM prompt. To obtain better results, we develop a novel schema linking algorithm. We find that using our schema linking algorithm provides a significant boost in accuracy scores. We also find that using the profile-generated metadata provides better results than using just the SME metadata supplied in the benchmark! Using fused metadata provides the best results, and the combination of techniques let us achieve the #1 spot on the BIRD leaderboard twice.

The BIRD query set and schemas are relatively simplistic, but interesting results can still be extracted. By using query log analysis, we can find a significant number (25% of total) of undocumented join paths. We can also find complex join predicates, and business logic for predicates and fields that are only documented in the oracles, or not at all.

We also investigate the use of the LLM SQL-to-text generation to create few-shot examples – a task made possible by the introduction of LLMs. While SQL-to-text has been used in e.g. [PLSC+24], we make an experimental study to evaluate the technique and find that the LLM can generate questions as good as or better than the human annotations.

# 2 Profiling

Database profiling has a huge literature dating back decades [AGN15]. The common idea is to analyze database contents to extract properties that aid in understanding database contents. Basic profiling takes a pass over a table and collects statistics such as

- The number of records in a table.
- For a field, the number of NULL vs. non-NULL values.
- For a field, the number of distinct values.
- For a field, the "shape" of a field, e.g. min and max, number of characters (digits), alphabet (upper/lower/punctuation/…), common prefixes, etc.
- For each field, a sample of the top-k field values.
- For each field, a minhash sketch.

Count distinct, and the set of top-k field values and their counts can be computed by approximate means [FFGM07] [IBS08] and these functions are increasingly present in commercial databases[3][4].

[1] A standard mobility equipment identifier. https://en.wikipedia.org/wiki/International_Mobile_Equipment_Identity.

[2] The BIRD question set includes an "oracle", which is a hint about how to write the query, specialized for each query.

A *minhash sketch* [B97] is a collection of K values computed by

$$m_i(f) = \min(h_i(v_j) \mid \text{over all values } v_j \text{ of field f})$$

for i ranging from 1 to K, and each $h_i$ is a different hash function.

The minhash sketch can be used to compute the *resemblance* of two contents of fields F and G, which is

$$res(F, G) = |F \cap G|/|F \cup G|$$

Given two minhash sketches m(f) and m(g), the resemblance between the values of fields f and g can be approximated by

$$res(f, g) = sum( if(mi(f)=mi(g), 1, 0) ), i \text{ in } 1, \ldots, K)/K$$

Given the minhash sketch of field f, the collection of fields g with a large intersection can be quickly computed. These can be used for tasks such as

- Finding join paths
- Imputing metadata from field f to field g.

Other more complex profiles can be collected, such as multi-field keys, functional dependencies and other kinds of dependencies [AGN15]. In this study we restrict ourselves to the basic profiles.

## 2.1 Using Profiling Information for Text-to-SQL

Actually using profiling information for text-to-SQL requires transforming the raw statistics into a form that the LLM can readily use. We describe the process using examples from BIRD.

We can start with `frpm.CDSCode`. A mechanically generated English language description of the profile for this field is:

> Column CDSCode has 0 NULL values out of 9986 records. There are 9986 distinct values. The minimum value is '01100170109835' and the maximum value is '58727695838305'. Most common non-NULL column values are '01100170109835', '01100170112607', '01100170118489', '01100170123968', '01100170124172', '01100170125567', '01100170130401', '01100170130419', '01100176001788', '01100176002000'. The values are always 14 characters long. Every column value looks like a number.

The next step is to use the English-language profile, the provided metadata for this field (which is just "CDSCode"), the table name, and the names of other fields in the table, to ask the LLM for a short description of the contents of the field. The resulting short description is:

> The CDSCode column stores unique 14-character numeric identifiers for each school in the database, where CDS stands for County-District-School.

The short description of `CDSCode` describes the format of the values in the field, and identifies its meaning: County-District-School. The LLM is able to pick up on the meaning of CDS because CDS is a common acronym of County-District-School. However CDS can also mean Cadmium Sulfide, credit default swap, counterfeit deterrence system, cross domain solution, and so on. But in the context of the table name (FRPM, or Free or Reduced Price Meal) and column names such as "Academic Year", "County Code", and so on, the most likely meaning of CDS is the one chosen.

While this short description is good for identifying the meaning of the field, more detail about the field values can guide the text-to-SQL LLM to use proper literal values. A long description is:

> The CDSCode column stores unique 14-character numeric identifiers for each school in the database, where CDS stands for County-District-School.The CDSCode column contains 14-character numeric strings with no NULL values, 9986 distinct values, ranging from '01100170109835' to '58727695838305'; common values include '01100170109835', '01100170112607', '01100170118489', '01100170123968', '01100170124172', '01100170125567', '01100170130401', '01100170130419', '01100176001788', '01100176002000'.

For another example where the LLM can guide the choice of literals for constraints, consider `` frpm.`Academic Year` ``. The provided metadata is "Academic Year", with the field value format left vague. Even the short LLM description is specific about the field value format:

> The `Academic Year` column stores the academic year for each record in the format 'YYYY-YYYY'.

A particularly striking example is `cards.leadershipskills`. This field contains JSON data, but this is not indicated in the field metadata:

> A list of formats the card is legal to be a commander in

The LLM recognizes the format of the field contents and provides this short summary:

> The leadershipSkills column stores JSON-formatted data indicating the formats in which a card is legal to be used as a commander, such as Brawl, Commander, and Oathbreaker

# 3 Schema Linking with Profile Metadata

The examples in the previous section make clear that an LLM can very often generate excellent descriptions of field contents and meanings. However these descriptions, especially the long descriptions, can overflow the token limit for LLM systems [TPCM+24]. In addition, we have observed that in the presence of long prompts, the LLM will pick up on the material in the beginning and the end, but tend to ignore the part in the middle (also observed by [TPCM+24]). The long field descriptions generated from profiling and LLM summarization are too long to be provided for context if such a description is provided for every field of every table in a database.

*Schema linking* [TPCM+24] [DZGM+23] [QHWY+22] [LPKP24] [GWLS+23] refers to identifying which fields are relevant to generating an SQL query in response to a question. Some authors [FPZE+24] have found that schema linking improves text-to-SQL performance even when the schema fits into the prompt context. In CHESS [TPCM+24], the authors found that perfect schema linking significantly improves performance. While some authors [MAJM24] have expressed an opinion that schema linking is not necessary with newer LLMs with large prompt contexts, industrial databases can have hundreds of tables each with hundreds of fields, so in practice schema linking is a necessity.

[3] https://docs.databricks.com/en/sql/language-manual/sql-ref-functions-builtin-alpha.html

[4] https://docs.snowflake.com/en/sql-reference/functions-aggregation

In this section, we describe how we performed schema linking in our BIRD submission. The value of the two types of LLM profile summaries (short and long) should be clear: the short summary is used to help with schema linking while the long summary is used for generating SQL from text.

There are four common schema linking mechanisms [TPCM+24], which can be used in combination:

- Metadata similarity search: search a vector database for fields whose name and/or metadata are semantically similar to the question.
- Column filtering: For each field, ask the LLM if the field is relevant.
- Table selection: Give the LLM the full schema and ask it to identify relevant tables.
- Column selection: Give the LLM the full schema and ask it to identify relevant fields.

We tried these schema linking techniques but obtained unsatisfactory results. The authors of [QLLQ+24] have observed that LLMs have good performance at tasks for which they have been trained, but poor performance on tasks outside of these boundaries. This property is known as *task alignment*. Specifically, we have observed that LLMs are not good at direct schema linking – identifying which tables/columns are relevant to a question – but are good at generating SQL. So our schema linking method is focused on generating SQL queries and gathering the fields they reference.

Another consideration is that the literals used in the question can help to indicate the field that should be constrained [TPCM+24]. One step in the algorithm is to identify fields which can contain a literal, and if the field involved in the constraint is not among that set, ask the LLM to rephrase the generated SQL using one of the fields in the set.

Yet another consideration is that recall is better than precision – it is better to put too many fields in the prompt (within limits) than too few.

The general outline of our schema linking algorithm is:

- For several different variants of field collections and their metadata:
    - Ask the LLM to generate an SQL query based on the question and in the context of the metadata variant
    - Collect the fields and literals in the generated query
    - Adjust the query to try to use fields which contain the literals, if needed.
- Use the collection of all fields returned by the different variants as the schema linked to the question.

In the remainder of this section, we describe in more detail the methods we have found to be effective. The schema linking algorithm requires a preprocessing step to aid in finding fields that match a literal.

1. For every field *f*,
    a. Fetch *N* distinct values of the *f*, or as many distinct values exist.
    b. Compute a string similarity index on these values
    c. Attach the string similarity index to the field's entry in the profile.

We found that a Locality Sensitive Hash (LSH) index [5] on shingles[6] is effective for the field value index as it provides an approximate match on values, but does not do semantic similarity. For the BIRD benchmark, we used N=10000. By contrast, CHESS [TPCM+24] indexes *all* values of all fields for literal matching. This technique is not scalable outside of small benchmark databases, so we limit ourselves to a moderate size sample.

A second semantic similarity index is computed on the profile using FAISS[7] [JDJ19]. This index is on textual descriptions on the profile of a field (i.e., the long summary), which allows for the efficient search of fields likely to be relevant to a textual query.

A key part of the schema linking algorithm is the metadata context. We use the following terms:

- Focused schema: Given a question, this is the set of fields which are textually similar to the users question, based on the string similarity index on the fields. In addition, literals are extracted from the question, and additional fields which include that literal in their values (using the LSH indices in the profile) are added to the focused schema.
- Full schema: All fields in all tables.
- Minimal profile: describe the field using the short LLM summary.
- Maximal profile: describe the field using the long LLM summary.
- Full profile: describe the field using the SME-supplied metadata along with the maximal profile.

Input: profile *Profile*, vector database *Index*, textual question *Question*, int *MaxRetry*

1. Let *Fields* be a set of fields and Lits be a set of literals
2. For each of the following five cases of schema *Schema*: a) focused schema, minimal profile; b) focused schema, maximal profile; c) full schema, minimal profile; d) full schema, maximal profile; e) focused schema, full profile; do the following
    a. Use the LLM to generate an SQL query *Q* in response to *Question* and *Schema*.
    b. Let *FieldsQ* be the fields referenced in *Q* and let *LitsQ* be the literals in *Q*.
    c. Let *LitFieldsQ* and *MissingLits* be empty lists
    d. For each literal *l* in *LitsQ*,

[5] https://github.com/ekzhu/datasketch

[6] https://pypi.org/project/kshingle/0.1.0/

[7] https://github.com/facebookresearch/faiss

i. Use the LSH indices in the profile to identify the fields *Fieldsl* which contain *l* as a value.

ii. If no field *f* in *Fieldsl* is in *FieldsQ*,

1. Add *Fieldsl* to *LitFieldsQ*
2. Add l to *MissingLits*

e. If *LitFieldsQ* is not empty and the number of retries is less than *MaxRetry*

i. Let *AugmentedSchema* be the schema augmented with any fields in *LitFieldsQ* which are not in Schema

ii. Write a prompt which asks the LLM to revise the SQL query *Q* suggesting the use of fields which contain literals in *MissingLits*, resulting in revised SQL query *Q*.

iii. Repeat steps 2.b through 2.e

f. Add *FieldsQ* and *LitsQ* to *Fields* and *Lits*

3. Return *FieldsQ* as the set of fields for providing context.

## 3.1 Example

We work through a simplified example from the BIRD benchmark:

> From the California Schools dataset, Please list the zip code of all the charter schools in Fresno County Office of Education.

The schema variant we will use is the full minimal profile, a sample of which is below:

> Field frpm.CDSCode means: The CDSCode column stores unique 14-character numeric identifiers for each school in the database, where CDS stands for County-District-School. This field joins with schools.CDSCode.
>
> Field frpm.`Academic Year`means: The `Academic Year` column stores the academic year for each record in the format 'YYYY-YYYY'.
>
> Field from.`County Code` means: The `County Code` column stores 2-character codes representing different counties.
>
> …
>
> Field schools.LastUpdate means: The LastUpdate column stores the date when each record was last updated.

A prompt is prepared and sent to the LLM, which responds with

```
SELECT T2.Zip FROM frpm AS T1 INNER JOIN schools
AS T2 ON T1.CDSCode = T2.CDSCode WHERE
T1.`Charter School (Y/N)` = 1 AND T1.`Country
Name`= 'Fresno County Office of Education’
```

We extract the following fields: frpm.CDSCode, frpm.`County Name`, frpm.`Charter School (Y/N)`, schools.CDSCode, schools.Zip; and the following literals: 'Fresno County Office of Education’.

Using the LSH indices in the schemas, we find that the literal does not occur in any field in the generated query, but does occur in fields frpm.`District Name`, satscores.dname, and schools.District. Another prompt is generated which recommends the LLM to use one of these fields to match the literal. The revised SQL query is

```
SELECT T2.Zip FROM frpm AS T1 INNER JOIN schools
AS T2 ON T1.CDSCode = T2.CDSCode WHERE
T1.`Charter School (Y/N)` = 1 AND T2.District =
'Fresno County Office of Education'
```

Now all literals are matched to a field in the query, so this field list is returned.

# 4 BIRD Submission Details

Our BIRD submission which achieved the #1 spot on the BIRD leaderboard on Nov. 11, 2024 primarily uses the techniques of database profiling, LLM profile summarization, and schema linking described in sections 2, 2.1, and 3. In this section, we describe some additional details of the BIRD benchmark submission.

Our BIRD submission makes use of few-shot examples [NZZR+23][GWLS+23][PL22] taken from the train query set. We use a technique described by [LPKP24]. In every question, we use the LLM to replace names with placeholders. The masked questions are put in a vector database with a reference to the corresponding SQL. To find few-shot examples, we mask the input question and find the 8 most similar questions in the vector database. These 8 queries are used as the few-shot examples.

As in Chase [PLSC+24], we generate multiple candidate queries and select one as the answer. To introduce variety into the candidate set, we use two techniques:

- Changing the LLM randomization seed.
- Changing the prompt by randomizing the order of the (schema linking-reduced) schema fields.

Our BIRD submission generates three candidates. These three candidates are checked for the validity of the SQL by using SQLglot[8].

We then check for SQL constructions that are likely to indicate an incorrect response. Some of these are checks on possible SQL problems. For example, a NULL value is ordered before all other values. So we check to ensure a NOT NULL predicate on *f* if:

- If the output is in ascending order on field *f*.
- If the select list contains the aggregate min(*f*).

Other checks relate to the apparent preferences of the authors of the SQL queries. For example:

- Check if a min/max query used a nested subquery instead of an Order By.
- Check if a query performs string catenation on fields instead of returning the fields individually.

If a bad SQL construction is detected, the LLM is asked for a correction with up to three retries.

We use majority voting among the candidates to pick one to be the final answer. Each of the up to three candidates are executed, and their results converted to sets. If there is agreement among two of the candidates, one of them is chosen. Else, an answer is chosen among the candidates randomly.

## 4.1 Experiments

The methods described in Sections 3 and 4 got us to the #1 spot on the BIRD benchmark leaderboard (currently at the #3 spot at the time of writing). As our interest is on the efficacy of automatically generated metadata, we ran some experiments on how additional field metadata helps to improve accuracy. For these experiments, we used the schema linking and additional techniques described in this section and used the GPT-4o LLM. We ran the experiments on the MiniDev questions/SQL set (of 500 questions). Our results are below.

[8] https://github.com/tobymao/sqlglot

| No metadata, no hints | 49.8% |
|---|---|
| Bird metadata, no hints | 59.6 |
| Profiling metadata, no hints | 61.2 |
| Fused (bird and profiling) metadata, no hints | 63.2 |
| Bird metadata, hints | 72.0 |
| Profiling metadata, hints | 72.6 |
| Fused metadata, hints | 73.0 |

**Table 1. Effects of metadata and hints on accuracy.**

A first takeaway is that the most powerful metadata available in Bird are the hints, which in MiniDev are very clear and precise. However such hints are not available in practice. If one has such exact information about how to write a query, one might as well develop a canned query system.

Without hints (for which our submission is at the #1 spot among those without hints), the text-to-SQL generation does surprisingly well even without field metadata. This result reflects the descriptive field and table names in many of the tables. Using field metadata improves accuracy as expected, but using profiling metadata results in a bigger accuracy boost than using the Bird-supplied metadata does! However there are details in the Bird metadata that are missing in the profiling metadata, so the fused metadata naturally provides the best accuracy (with or without hints).

A next question to ask is, how well does schema linking work? We compare using the full schema, our schema linking (described in Section 3), and perfect schema linking. We used the fused metadata for these experiments. The results are in the table below.

| Full schema, no hints | 61.2% |
|---|---|
| Our schema linking, no hints | 63.2 |
| Perfect schema linking, no hints | 69.0 |
| Full schema, hints | 71.2 |
| Our schema linking, hints | 73.0 |
| Perfect schema linking, hints | 77.4 |

**Table 2. Effects of schema linking on accuracy.**

Recent papers [MAJM24][PLSC+24b] claim that schema linking is not needed when using frontier models such as GTP-4o. However these results show that this is not the case even for the small schemas in the BIRD benchmark. The full schema case corresponds to no-schema-linking. The algorithm described in Section 4 provides a significant improvement, so effective schema linking does help. However perfect schema linking provides a large jump in scores. Clearly, further research on schema linking for text-to-SQL is needed.

#### 4.1.1 Detailed Discussion

In this section, we describe some peculiarities we observed in these experiments. One phenomenon that we observed is that providing profile metadata can lead to generated queries being flagged as incorrect. For example, in Q356, the question is

> How many cards have infinite power?

With the hint

> infinite power refers to power = '*'

Profiling detects the presence of special symbols, and the LLM summary contains the phrase

> special symbols like "∞" for infinite power

The gold SQL is

```
SELECT COUNT(*) FROM cards WHERE power = '*'
```

But the predicate should be power='∞'.

Another example is Q1260, which has question

> Please list the ID of the patient whose RF is normal and who is older than 60.

With hint

> normal RF refers to RF < 20; don't have thrombosis refers to Thrombosis = '0'

The gold SQL is

```
SELECT COUNT(DISTINCT T1.ID) FROM Examination
AS T1 INNER JOIN Laboratory AS T2 ON T1.ID =
T2.ID WHERE T2.RF < 20 AND T1.Thrombosis = 0
```

This query is wrong is several ways, but we will focus on the predicate on Laboratory.RF. The hint states that the predicate to use is RF<20 and this predicate is pasted in. But RF is of type text, so the proper predicate should be `CAST(T2.RF AS REAL) < 20`.

In some cases, the query generated using the full schema is flagged as correct, but with the linked or perfect schema, its flagged as incorrect. One example is Q1480, where there is an issue in formatting yearmonth.Date with the linked and perfect schema, but not the full schema. Another example is Q1505, where the choice between returning count(*) vs. count(distinct CustomerID) depends on the length of the schema. Here the linked and perfect schema SQLs are correct but the gold SQL is not, so they are marked as wrong. These problems are due to the instability of LLM answers.

## 5 Query Log Analysis

Most DBMS systems collect query logs – a log of all the SQL query text submitted to the DBMS. Many of these queries are written by expert SMEs. So, by extracting features from the log, we can obtain SME information without the need for extensive interviews. In addition, query logs likely contain features that SMEs don't know about or have forgotten.

The query log is kept for some window (e.g. 90 days) and contains a variety of additional metadata such as the time of submission, user ID, query status (succeed/fail), performance metrics, or even the query plan. While these metadata fields are valuable for filtering, in the study we focus on the query text.

In some data analytics platforms, e.g. DataBricks, a large portion of the queries submitted to the system do not have textual SQL queries. This is the case when data analytics frameworks such as PySpark[9] are used. Pyspark primarily operates on *dataframes*, (which correspond to Spark RDDs) using a sequence of dataframe manipulation methods to construct an execution plan of select, project, join, aggregation, etc. operations. These are collected until an action triggers the dataframe evaluation. The execution plan is optimized and evaluated, generally in the same way that an explicit SQL query would be. So, dataframes allow the construction of SQL-equivalent plans but no SQL text is involved and no SQL text is logged.

[9] https://spark.apache.org/docs/latest/api/python/index.html

However Databricks keeps logs of query plans, which can be handled in a manner similar to that of textual SQL queries. We will note how query plans can be handled in the discussion of query log processing. Pyspark is becoming popular in other databases, such as e.g. Snowpark[10] in Snowflake.

## 5.1 Query Log Processing

The raw query text is not directly useful in this section, instead it has to be processed to extract interesting features. To be useful for SQL generation, all fields referenced in the query must be traced back to their source table. Further, in the presence of subqueries, the formula used to compute a field should be substituted for that field.

For example, consider the following query:

```
Select A.uop_cd, A.trans_amt+P.total_planned as
current_spend_and_planned
From accounting_table A, (
  Select source_code, sum(planned_amt)as
total_planned
  From planning_table
  Where country_code='USA'
  Group By source_code) P
Where A.uop_cd=P.source_code
```

In the select clause, A.uop_cd can be resolved to be accounting_table.uop_cd. The second element has an element sourced from a subquery. Resolving the formula for P.total_planned, we can determine that output field current_spend_and_planned is sourced from

```
accounting_table.trans_amt +
sum(planning_table.planned_amt)
```

To perform field resolution and feature extraction, the first step is to use an SQL parser to convert the SQL statement into an Abstract Syntax Tree (AST). There are many open source SQL parsers, e.g. sqlglot[11] and Jsqlparser[12]. We developed our own using the Python Lark[13] parser.

To simplify the discussion, we assume that the AST returns a node which represents a regular query, e.g. select/from/where/group-by/having. The only subqueries (used for field resolution) are in the From clause. The goal of the algorithm is to resolve the formulas used to compute the fields of the subqueries, if any, to use for resolving fields in the top-level query. The actual algorithm we use has a variety of complexities to handle those of SQL, but they are not important here.

The main data structure is a *subquery summary* table, which maps a field returned by a subquery with the formula which computes it. For example, the subquery summary for P is

| **Field** | **Formula** |
|---|---|
| source_code | planning_table.source_code |
| total_planned | Sum(planning_table.planned_amt) |

[10] https://docs.snowflake.com/en/developer-guide/snowpark/index

[11] https://github.com/tobymao/sqlglot

[12] https://github.com/JSQLParser/JSqlParser

[13] https://github.com/lark-parser/lark

The algorithm for field resolution is:

Resolve_fields(*root*)

1. For every subquery *q* in the From clause
    a. Summary(*q*) = resolve_fields(*q*)
2. *Query_summary* = <empty table>
3. For every field *f* in the Select list
    a. Resolve the fields referenced in *f*'s formula using the tables and subquery summaries in the From clause
    b. *Query_summary(f)* = <resolved formula>
4. Return *Query_summary*

For example, to resolve the top level in the example query, we process each returned field in turn. `Uop_cd` is computed from `A.uop_cd`, which resolves to `accounting_table.uop_cd`. Current_spend_and_planned is computed from `A.trans_amt+P.total_planned`. The two fields in this formula resolve to `accounting_table.trans_amt` and `sum(planning_table.planned_amt)`. So spend_and_planned is computed from their sum, with result:

| **Field** | **Formula** |
|---|---|
| Uop_cd | `Accounting_table.uop_cd` |
| Current_spend_and_planned | `accounting_table.trans_amt + sum(planning_table.planned_amt)` |

## 5.2 Query Plan Processing

Ideally, one has access to query logs containing the SQL text. However in some cases, only a query plan is available. We encountered this issue when trying to understand queries in DataBricks. Many data science workloads do not submit textual SQL queries, but rather are written in a programming system such as PySpark. For example, the query

```
Select a, b from Table where c=5
```

Could be expressed as

```
Df =
spark.read.format("delta").load(<Table_delta_f
ile>). where(col('c')=5).select('a', 'b')
```

These constructions are not rendered into SQL, instead they create a query plan which is optimized and executed. Given that Spark integration is a desirable feature, additional vendors are adding similar execution environments - e.g. Snowpark/Snowflake.

The simple query above would be rendered into a query plan in the expected way: a table scan, followed by a Select operator, followed by a Project operator, followed by an Output operator[14]. Each operator can be viewed as a separate subquery, so the algorithm for resolving field formulas through subqueries applies to well-annotated query plans also.

[14] Simple selection and filters will get pushed into the scan after optimization.

CHASE-SQL [PLSC+24] processes the SQLite query plan to extract features, to ensure diversity in SQL queries generated in response to a question, for candidate selection.

## 5.3 Feature Extraction

We can extract further features from this query, for example

- Spend_and_planned is computed from `accounting_table.trans_amt + sum(planning_table.planned_amt)`
- There is a constraint `planning_table.country_code='USA'`, or alternatively `planning_table.country_code=`<string>
- `Planning_table.source_code` is used as a group-by variable
- There is a join path `accounting_table.uop_cd=planning_table.source_code`

These features provide valuable metadata about the use of accounting_table and planning_table. The named field, spend_and_planned, names the formula in bullet point 1. So if a question asks current and planned spending, textual similarity leads to the formula. The labeling of a join path from accounting_table to planning_table provides solid information about how to perform joins if primary key/foreign key information is missing from the schema (as it often is).

The types of features one can extract from a query include:

- Named select fields
- Unnamed select fields (i.e. no AS clause, or a trivial AS clause)
- Non-join field constraints – i.e. between fields of the same range variable
- Join field constraints – between fields of different range variables
- Constraints in the ON clause.
- Sets of constraints in the ON clause – two or more field constraints might be needed for the join from R to S.
- Group-by formulas
- Group-by sets – all fields (formulas) in a group-by clause
- WITH subqueries. The naming of a subquery often indicates its purpose.
- Sets of tables referenced
- Set of fields referenced
- Query complexity

Query log features help in a variety of ways. For one, primary/foreign key constraints are often not well documented, and many join constraints involve multiple fields and data transformations. For another, extracted features can show details of how a database should be queried. A common strategy in text-to-SQL is the use of *few-shot examples* [NZZR+23] [GWLS+23][PL22] – sample queries generally found by semantic similarity of their associated questions to the target question. However, the few shot queries might not contain the necessary details, and might have the wrong query "shape". Further, features can be expressed with less text that few-shot examples, reducing prompt size.

## 5.4 Experiments

BIRD does not reveal any information about test database used for ranking, so no query logs are available for analysis. Instead, we use BIRD as a test suite for determining if we can use log analysis for detecting missing metadata. We use the BIRD dev test suite from Sept 2023 (dev 9-23), as well as the newer cleaned-up version of June 2024 (dev 6-24), as the unrevised version is likelier to reflect actual industrial databases.

Our experiments focus on three questions:

1. Can we find equality (e.g. pk-fk) join constraints that are not documented in the SQLite schema?
2. Can we find other interesting join constraints?
3. Can we find interesting named formulas (business logic).

For interesting constraints and named formulas, we compare the features that we find to the provided metadata, and also to the hints associated with each query. We note that in an industrial application, no hint is available, so one must use few shot examples or relevant query features.

### 5.4.1 Primary key - Foreign key / Equality Constraints

Using the dev queries as the query log, we extract all constraints of the form R.f=S.g. This yields 109 distinct constraints for dev 9-23 and 107 for dev 6-24. Comparing the constraint lists, there are 3 new constraints and 5 missing constraints in dev 6-24 as compared to dev 9-23, reflecting a revision of the gold standard queries. An example of an added constraint is

card_games.legalities.format=card_games.maxbanned.format

and an example of a missing constraint is

financial.disp.account_id=financial.trans.account_id

Extracting the PK-FK constraints from the BIRD documentation required some manual intervention. We use the SQLite schemas of the SQLite databases to extract the foreign keys using `PRAGMA foreign_key_list`. On processing the results, we found that three constraints were improperly specified (for both):

debit_card_specializing.yearmonth.CustomerID = debit_card_specializing.customers.CustomerID
european_football_2.Match.league_id = european_football_2.League.id
european_football_2.Match.country_id = european_football_2.Country.id

The problem being that the foreign key field was not specified in the SQLite schema, and the sqlite pragma wasn't able to identify the referenced field. So, the field on the right hand side was empty. We filled these in by hand. The result was 109 constraints for dev 9-23 and 104 for dev 6-24 (the unused databases card_games_2 in dev 9-23 was removed for dev 6-24).

We normalized the two sets of constraints and took set differences to find how the results differed. From the SQLite dev 9-23 constraints, there were 29 equality constraints that were never used, and from dev 6-24 there are 24 unused constraints. Examples include:

card_games_2.foreign_data.uuid = card_games_2.cards.uuid
european_football_2.Match.away_player_11 = european_football_2.Player.player_api_id
toxicology.connected.atom_id = toxicology.atom.atom_id

From the constraints extracted from log analysis, there are 29 constraints detected in dev 9-23 queries but not documented in the SQLite schema, and for dev 6-24 there are 27. Examples include:

debit_card_specializing .gasstations.gasstationid = debit_card_specializing .transactions_1k.gasstationid
card_games.cards.setcode = card_games.sets.code
thrombosis_prediction .examination.id = thrombosis_prediction .laboratory.id

Of the discovered constraints, 4 equated different field names (e.g. cards.setcode=sets.code). If the hand-constructed constraints are excluded, there are 32 / 30 discovered constraints of which 6 equate different field names. So, 27% / 25% of the field-to-field equality constraints actually used in a dev query are either undocumented or require hand extraction. While dev 6/24 has slightly fewer undocumented equality constraints, a large fraction of these constraints are undocumented in the schema.

Many of the missing equality constraints have either the same field name, or they both end in "id". So the LLM can make a good guess about how to perform a join. However there are two problems. First, industrial databases (especially those grown by accretion) often do not have such convenient naming systems. Second, its better to know the join path than to need to guess it.

For an experiment, we counted the number of fields per table that end in "id". We found that there are an average of 1.8 id fields per table, and a maximum of 15 (in cardgames.cards). 20 tables have no id field, and 23 have one id field. So guessing an equality predicate in BIRD can still present challenges.

### 5.4.2 Other Join Constraints

Many more join constraints can be found (examples from dev 6-24). For example, there are two multi-field joins:

Q782: colour.id=superhero.eye_colour_id And colour.id=superhero.hair_colour_id

Q1016: results.raceid=pitstops.raceid And results.driverid=pitstops.driverid

None of these 4 individual predicates are listed in the SQLite schema, although the hint in 782 suggests the join. We can also find joins that require computation:

Q234: "bond.molecule_id | |'_1'=connected.atom_id And bond.molecule_id || '_2'=connected.atom_id2 And bond.bond_id=connected.bond_id

The metadata for the bond.molecule_id field does mention the encoding standard. However no join path from bond.molecule_id to connected.atom_id is explicitly documented. There are also many multi-table join predicates:

Q146: card.disp_id=disp.disp_id AND client.client_id=disp.client_id AND disp.account_id=loan.account_id

These constraints indicate join patterns, which might go through an otherwise unrelated linkage table.

When a table has multiple date fields, understanding which should be used in a constraint can be obscure. For example, california_schools.schools has three date fields, opendate, closedate, and lastupdate. Opendate is constrained in 6 queries (4, 27,39, 47, 66, 87) and closeddate is constrained in three (27, 67, 68), while lastupdate is never constrained.

Finally, we can find date constraints:

strftime('%Y', california_schools.schools.opendate)='1980'
california_schools.schools.opendate>'2000-01-01'

### 5.4.3 Interesting Formulas

In our experience, queries contain many named non-trivial formulas (capturing useful business logic). The queries in BIRD generally don't name elements in the select clause, but there are some. For example, Q221 has a pair of named formulas in the Select clause, which we paraphrase as:

atom_id1 is computed from substr(bond.bond_id, 1, 7)
atom_id2 is computed from (bond.molecule_id)||(substr(bond.bond_id, 8, 2))

The field metadata contains the following:

bond_id: unique id representing bonds; TRxxx_A1_A2: TRXXX refers to which molecule A1 and A2 refers to which atom
molecule_id:identifying the molecule in which the bond appears

So there is an indication of the structure of bond_id but the formulas used in Q221 are not clear from the text. Some other examples are:

Q1499: monthlyconsumption is computed from (sum(debit_card_specializing.yearmonth.consumption))/(12)

Q215: iodine_nums is computed from count(DISTINCT case When ('i'=toxicology.atom.element) Then toxicology.atom.atom_id Else Null End)

Q222: diff_car_notcar is computed from (count(case When ('+'=toxicology.molecule.label) Then toxicology.molecule.molecule_id Else Null End))-(count(case When ('-'=toxicology.molecule.label) Then toxicology.molecule.molecule_id Else Null End))

In each of these cases, the "evidence" hint suggests the use of these formulas. However, in practice these hints are not available. But they can be extracted by query log analysis.

### 5.4.4 Summary

We have shown that many useful query features can be found in the collection of dev queries, considered as a query log. For example, 25% of the equality joins that are used in at least one query are not documented in the SQLite schema. However, the usefulness of this information is limited in the context of the BIRD benchmark, which is fairly simple and readable. Field names are highly suggestive of e.g. join paths, and few formulas are explicitly named in the gold queries using AS (correctness checking is done using field positions, not names). The hints provided with the question generally contain the query features to be used in the generated query – a very unrealistic situation. A better test would list query features and require the submission to do query feature linking.

## 6 Sql to Text

The use of question/SQL pairs as few-shot examples has been shown to be an effective means of boosting text-to-SQL performance examples [NZZR+23][GWLS+23][PL22] and has been used in our BIRD submission, as described in section 4. However, generating these pairs creates a very large workload for the SMEs who must think up questions and write the corresponding SQL queries. For the Spider benchmark, [YZYY+18], students spent 1,000 hours creating 5700 question/answer pairs. The BIRD benchmark does not list the number of work hours, but states that 11 contributors and three assessors were employed in developing 12000 question/answer pairs, and that $98,000 was spent.

If one has access to query logs, then one can sample a selection of these queries and use an LLM to generate a corresponding question from them. The procedure we use is:

Input: SQL query *Q*, full schema *S*

1. Analyze *Q* to determine the set of fields *F* referenced in *Q*.
2. Extract a focused schema *FS* by selecting from *S* the fields in *F*
3. With the context of the focused schema *FS*, ask the LLM to create a long question and a short question from query *Q*.

That is, by query analysis, one can obtain perfect schema linking [TPCM+24] – helping to make SQL-to-text an easier problem than text-to-SQL.

Query logs can contain a very large number of distinct queries, but only an interesting and representative sample should be selected for few-shot examples. For example, one might be focused on generating SQL for a subset of tables, or one might choose examples which use an interesting formula such as "(bond.molecule_id)||(substr(bond.bond_id, 8, 2))". We have found the following procedure to be useful.

1. For each query *Q* in query log *L*, analyze *Q*, extract its features, and associate them with *Q* (e.g. in a JSON file).
2. Provide the summary of extracted features to an SME, who identifies features *F*, containing a collection of query features to match against. Aggregate the collection of features into a set of features *FS*.
3. For each set of features *fs* in *FS*, collect the list set of queries *Qfs* that contain *fs*.
4. Return the union of all *Qfs*.

## 6.1 Experiments

Our experiments consist of generating text from SQL, and then comparing the generated questions to the supplied question and SQL. We used BIRD Minidev [15] as the source of the question/SQL pairs. However there are 500 question/SQL pairs in Minidev, the grading process is very time consuming, and we have limited human resources. So we further selected every 6th entry (i.e., the first, 7th, 13th, etc. entry) for use in evaluation.

The question/SQL pairs were human-generated in the text-to-SQL direction. For our experiments, we treat the corpus as having been generated in the opposite direction: given an SQL query, what is it asking?

We generate both the long and short question, and use the best result for grading – which we consider reasonable both for query explanation and for query retrieval from a vector database for few-shot selection.

Our grading is subjective, so list the SQL, supplied question, generated questions and their ratings in the Appendix. Our ratings are:

- Bad : The question is not related to the SQL.
- Bad+ : The question is related to the SQL, but misses important details.
- Good-: The question generally matches the SQL, but misses some important detail.
- Good : The question matches the SQL and is on par with the supplied question in terms of accuracy and readability.
- Good+ : The question matches the SQL and is better than the supplied question.

We note that good+ includes entries in which the supplied question and the SQL are not in agreement (problems with question/SQL pairs have been previously noted [TPCM+24]). We use four different kinds of metadata:

- Base : no metadata.
- Bird : The benchmark-supplied metadata (i.e., from the *.csv files in the database_description subdirectories).
- LLM : The short LLM summary generated from the field profile.
- Fused: Both the Bird and the LLM metadata.

Our results, across all difficulty levels are below. The results sliced on difficulty level (simple/moderate/challenging) are similar:

| Rating | Base | Bird only | LLM only | Fused |
|---|---|---|---|---|
| Bad | 0 | 0 | 0 | 0 |
| Bad+ | 1 | 1 | 3 | 0 |
| Good- | 16 | 14 | 12 | 2 |
| Good | 53 | 55 | 55 | 68 |
| Good+ | 13 | 14 | 13 | 13 |

**Table 3. SQL-to-text evaluation.**

Even with no metadata, the SQL-to-text performance is surprisingly good – almost as good as human annotation. With fused metadata, the generated questions are significantly better than the human annotation. We conclude that by using the techniques described in Section 2, query extraction from a query log plus SQL-to-text generation is an effective technique for generating few-shot examples.

The human-generated question is worse in 13 out of 83 sample questions (16%). This is likely due to the tedious and exhausting nature of generating 12000+ question/SQL pairs. In the remainder of this section, we explore some examples.

We start with the example where the generated question is rated "bad+" (question_id 93). The SQL is

```
SELECT COUNT(T1.client_id) FROM client AS T1
INNER JOIN district AS T2 ON T1.district_id =
T2.district_id WHERE T1.gender = 'M' AND T2.A3 =
'north Bohemia' AND T2.A11 > 8000
```

The bad+ question using LLM metadata is

> How many men from districts in north Bohemia with populations over 8000 are clients?

Field A11 refers to salary not population so the generated question is significantly off. A good- base-generated question is:

[15] https://github.com/bird-bench/mini_dev

> How many men from the North Bohemia district with an A11 value over 8000 are clients?

This question also misses the meaning of field A11, but does not try to guess the meaning. The question with fused metadata indicates a salary of 8000 or more, which is correct.

An example entry where all of the generated questions are good is question_id 710. The SQL is

> SELECT COUNT(T1.id) FROM comments AS T1 INNER JOIN posts AS T2 ON T1.PostId = T2.Id WHERE T2.CommentCount = 1 AND T2.Score = 0

While the supplied question is

> In posts with 1 comment, how many of the comments have 0 score?

The generated questions are similar, though the one generated with the fused metadata is more accurate:

> How many comments are linked to posts with only one comment and no upvotes?

In 8 of the 83 total questions, all of the generated questions are rated “good”, but we added a clarification note, generally indicating that the supplied question is vague or has poor grammar. An example is question_id 39 with SQL:

> SELECT AVG(T1.NumTstTakr) FROM satscores AS T1 INNER JOIN schools AS T2 ON T1.cds = T2.CDSCode WHERE strftime('%Y', T2.OpenDate) = '1980' AND T2.County = 'Fresno'

And supplied question

> What is the average number of test takers from Fresno schools that opened between 1/1/1980 and 12/31/1980?

Even the question generated with the base metadata is more accurate:

> What's the average SAT participation for schools opened in 1980 in Fresno County?

For some generated questions labeled good+, an example where all generated questions are good+ is question_id 112, with SQL

> SELECT T1.A2 FROM district AS T1 INNER JOIN client AS T2 ON T1.district_id = T2.district_id WHERE T2.birth_date = '1976-01-29' AND T2.gender = 'F'

The supplied question is

> For the female client who was born in 1976/1/29, which district did she opened her account?

However, there is nothing in the SQL which suggests that there is only one match. A more accurate question is:

> Which districts have female clients born on 29th January 1976?

An example of a supplied question which can be considered accurate, but which have poor grammar is question_id 862:

> For the Bahrain Grand Prix in 2007, how many drivers not finished the game?

An example where the supplied question does not match the SQL is question_id 231, with SQL

> SELECT T.bond_type FROM ( SELECT T1.bond_type, COUNT(T1.molecule_id) FROM bond AS T1 WHERE T1.molecule_id = 'TR010' GROUP BY T1.bond_type ORDER BY COUNT(T1.molecule_id) DESC LIMIT 1 ) AS T

And supplied question

> Which bond type accounted for the majority of the bonds found in molecule TR010 and state whether or not this molecule is carcinogenic?

The select list has no indication of carcinogenic status.

## 7 Related Work

Text-to-SQL code generation has attracted a great deal of attention in recent years [CLHY+22][QHWY+22][YWDD17][XLS17][ZM96]. Development has been accelerated by the release of standardized benchmarks: WikiSQL [16] [ZXS17], Spider[17] [YZYY+18] and BIRD[18] [JHQY+23]. In this discussion, all scores and rankings are at the time of writing (Jan. 2025).

In [GWLS+23], the authors put a corpus of Spider questions in a vector database and extracted few-shot examples by similarity to the posed question. This technique and a tuned LLM got them the #1 spot on the Spider leaderboard (currently at #2).

The CHESS[19] submission to BIRD [TPCM+24] uses LLM-driven schema linking techniques: column filtering, table filtering, and column selection. This plus query revision and query candidate selection got them on the #1 spot (currently at #7).

The BIRD submission from Distillery [MAJM24] takes a different approach. They argue that newer LLMs remove the need to do schema linking. They achieved the #1 spot on BIRD, and are currently at #6.

IBM achieved the #1 spot on Bird, and is currently at #4 spot. They have not posted a paper but marketing materials[20] state that they use “extractive schema-linking” and a tuned version of their Granite LLM.

Chase [PLSC+24] has the current #2 spot on the BIRD benchmark. Their paper describes a number of interesting techniques, including methods for generating diverse query candidates. They use query plan analysis to determine the “shape” of the query, and try to get candidates with a variety of query shapes. They also use a tuned version of the Google Gemini LLM. Our submission and the Google submission have been trading the #1 spot on the Bird benchmark, both recently beaten by Alibaba.

XiYan-SQL [GLLS+24] has the current number spot on the BIRD benchmark. Among the techniques used is to generate candidate SQL using several different models, and then use a fine-tuned selector model to pick the result.

We explore the use of well-known, and also newer, metadata extraction techniques for text-to-SQL generation. Database profiling [AGN15] has a large literature. Query parsing and feature extraction is at the core of query planning, and has been used to e.g. find join paths in large, complex databases [YPS09]. A new technique is the conversion of SQL to text questions. This technique has been used for candidate selection by Chase [PLSC+24].

## 8 Conclusions

[16] https://www.kaggle.com/datasets/shahrukhkhan/wikisql

[17] https://yale-lily.github.io/spider

[18] https://bird-bench.github.io/

[19] https://github.com/ShayanTalaei/CHESS

[20] https://research.ibm.com/blog/granite-LLM-text-to-SQL

In our decades of experience developing queries on large and complex industrial databases, we have found that the hardest task is understanding the database contents. After that, writing SQL queries is generally straightforward. With that motivation, we develop several strategies for automatic metadata extraction for text-to-SQL generation: database profiling, query log analysis, and SQL-to-text generation.

Database profiling [AGN15] is a collection of well-known techniques used to characterize database contents. We combine profiling results with basic table metadata, and find that an LLM can provide a surprisingly good summary of the field meaning. In one example, the summary discovered the JSON format of the field contents.

For evaluation, we developed a novel schema linking strategy that relies on task alignment. Using this schema linking strategy, we find that field metadata generation using profiling and LLM summarization is highly effective. Using BIRD as a test suite, we find that using profiling metadata leads to higher accuracy than using the supplied metadata (with or without hints)! Using fused metadata unsurprisingly provides the highest accuracy. Our BIRD benchmark submission used this technique and achieved the #1 position on the leaderboard (with hints) from Sept. 1 to Sept. 23, 2024 and from Nov. 11 to Nov 24, 2024, and retains the highest position on the leaderboard without hints at the time of the writing.

We applied query log analysis to the BIRD dev query set. Even though primary key – foreign key dependences are well documented in the schema, the log analysis was able to find 25% more useful equality constraints than are documented. Query log mining also found multi-field join constraints, join constraints that involve computations, field constraint formulas, and formulas in the select clause. Many of these features are either documented in the hints (which are not present in actual text-to-SQL tasks) or not documented.

Finally, we ran experiments on SQL-to-text, which can be used for query explanation or for generating few-shot examples from query logs. Even with no metadata, the LLM scored nearly as well as the human-generated questions. This result reflects the clean and clear naming in most parts of the BIRD schemas. With fused BIRD/profiling metadata, the generated questions scored significantly better than the human generated questions. This is likely due to the tedious nature of question/SQL generation for 12000+ query pairs in dev and test alone, leading to frequent human mistakes.

## Suggestions for Future Benchmarks

The BIRD benchmark is a significant step up in complexity as compared to the Spider benchmark. This can be seen in the accuracy scores of the entries at the #1 spot: 91.2% vs. 74.79%. Part of the increase in difficulty is in the ambiguity of the schema and field names, and the sometimes-poor field metadata. These issues allow our automatic metadata extraction techniques to work well.

However the schemas are still simple and have highly suggestive names and join paths. This can be seen in the good performance of text-to-SQL and SQL-to-text without any field metadata.

In our experience, industrial databases are far more complex and dirty. Field and table names tend to be obscure, join paths are not obvious from the field names, and documentation tends to be incomplete and out of date. Many databases contain thousands of tables, many of which have 500+ fields. Generating SQL from questions for these databases requires a significant step up from the schema linking techniques developed to date.

BIRD relies on per-question hints to supply special information, (e.g. formulas, predicates, join paths) to allow that query-generating prompt to supply these non-obvious details. However per-query hints are very unrealistic in real-world databases. A more realistic benchmark would provide a table of query features which apply to the database in question. This change would allow for the testing of query feature linking as well as schema linking.

The BIRD benchmark has a second leaderboard for query execution efficiency (our submission is at the #3 spot). This is a curious metric for rating text-to-SQL systems, as the results are very highly dependent on the query optimizer, which varies from DBMS to DBMS, and is frequently updated in actively supported DBMSs.

However, it is possible to engineer performance bombs in the data, which a good text-to-SQL system should be able to detect, or at the least warn about. One common example is the presence of default values, instead of NULL values. Often there are multiple default values. For example, a telephone_number fields might have 10% of its entries with value 123-456-7890, and another 5% with value 999-999-9999. If one does a join on telephone_number and both fields have a large overlap in their default values (likely), then a join which should return millions of records will return trillions of records and most of these records are garbage.

## ACKNOWLEDGMENTS

We'd like to thank Chris Vo and Pavan Tagarisa for their help with creating the BIRD benchmark submissions, and the BIRD team for their prompt evaluations.

## Appendix

In this section, we show the SQL, supplied question, generated questions and their ratings from the experiments in Section 6.

Question 1471:

SQL:

```
SELECT CAST(SUM(IIF(Currency = 'EUR', 1, 0)) AS FLOAT) / SUM(IIF(Currency = 'CZK', 1, 0)) AS ratio FROM customers
```

Supplied question: What is the ratio of customers who pay in EUR against customers who pay in CZK?

Base generated questions: Q1:: What is the ratio of the number of customers using the EUR currency to the number of customers using the CZK currency in the customers table? Q2:: What is the EUR to CZK customer ratio?

Base question rating: good.

Bird only generated questions: Q1:: What is the ratio of the total number of customers using the Euro (EUR) to the total number of customers using the Czech koruna (CZK) in the customers table? Q2:: What is the EUR to CZK customer ratio?

Bird only question rating: good+. Reason: Q2 is succinct and expands the acronyms.

LLM only generated questions: Q1:: What is the ratio of the total number of customers using the EUR currency to the total number of customers using the CZK currency in the customers table? Q2:: What is the EUR to CZK customer ratio?

LLM only question rating: good.

Fused generated questions: Q1:: What is the ratio of the total number of customers using the EUR currency to the total number of customers using the CZK currency in the customers table? Q2:: What is the EUR to CZK customer ratio?

Fused question rating: good.

---------------------------------------------------

Question 1481:

SQL:

```
SELECT CAST(SUM(IIF(T1.Segment = 'SME', T2.Consumption, 0)) AS REAL) / COUNT(T1.CustomerID) - CAST(SUM(IIF(T1.Segment = 'LAM', T2.Consumption, 0)) AS REAL) / COUNT(T1.CustomerID) , CAST(SUM(IIF(T1.Segment = 'LAM', T2.Consumption, 0)) AS REAL) / COUNT(T1.CustomerID) - CAST(SUM(IIF(T1.Segment = 'KAM', T2.Consumption, 0)) AS REAL) / COUNT(T1.CustomerID) , CAST(SUM(IIF(T1.Segment = 'KAM', T2.Consumption, 0)) AS REAL) / COUNT(T1.CustomerID) - CAST(SUM(IIF(T1.Segment = 'SME', T2.Consumption, 0)) AS REAL) / COUNT(T1.CustomerID) FROM customers AS T1 INNER JOIN yearmonth AS T2 ON T1.CustomerID = T2.CustomerID WHERE T1.Currency = 'CZK' AND T2.Consumption = ( SELECT MIN(Consumption) FROM yearmonth ) AND T2.Date BETWEEN 201301 AND 201312
```

Supplied question: What is the difference in the annual average consumption of the customers with the least amount of consumption paid in CZK for 2013 between SME and LAM, LAM and KAM, and KAM and SME?

Base generated questions: Q1:: What is the difference in average minimum monthly consumption between 'SME' and 'LAM' segments, 'LAM' and 'KAM' segments, and 'KAM' and 'SME' segments for customers with the 'CZK' currency within the year 2013? Q2:: What are the average differences in the least monthly consumption for 'SME', 'LAM', and 'KAM' customer segments using Czech koruna in 2013?

Base question rating: good. Reason: Q2.

Bird only generated questions: Q1:: What is the difference in average consumption per customer between the SME and LAM segments, LAM and KAM segments, and KAM and SME segments for customers with the currency 'CZK' who had the minimum consumption in the year 2013? Q2:: What are the average consumption differences per customer among SME, LAM, and KAM segments for 'CZK' currency customers with the least consumption in 2013?

Bird only question rating: good.

LLM only generated questions: Q1:: What is the average difference in the minimum consumption between the 'SME' and 'LAM' segments, the 'LAM' and 'KAM' segments, and the 'KAM' and 'SME' segments for customers with the currency 'CZK' within the year 2013? Q2:: What are the average differences in the least consumption among 'SME', 'LAM', and 'KAM' customer segments in CZK for 2013?

LLM only question rating: good.

Fused generated questions: Q1:: What is the difference in average minimum monthly consumption in the year 2013 between the SME and LAM segments, between the LAM and KAM segments, and between the KAM and SME segments for customers using the CZK currency? Q2:: What is the average minimum consumption difference in 2013 between customer segments SME-LAM, LAM-KAM, and KAM-SME using Czech currency?

Fused question rating: good.

---------------------------------------------------

Question 1493:

SQL:

SELECT CAST(SUM(IIF(Consumption > 528.3, 1, 0)) AS FLOAT) * 100 / COUNT(CustomerID) FROM yearmonth WHERE Date = '201202'

Supplied question: In February 2012, what percentage of customers consumed more than 528.3?

Base generated questions: Q1:: What is the percentage of customers with a consumption greater than 528.3 units in February 2012? Q2:: What percent of February 2012 customers used more than 528.3 units?

Base question rating: good.

Bird only generated questions: Q1:: What is the percentage of customers who had a consumption greater than 528.3 units in February 2012? Q2:: What percent of customers consumed more than 528.3 units in Feb 2012?

Bird only question rating: good+.

LLM only generated questions: Q1:: What is the percentage of customers whose consumption exceeded 528.3 in February 2012? Q2:: What percentage of customers had a consumption over 528.3 in February 2012?

LLM only question rating: good.

Fused generated questions: Q1:: What percentage of customers consumed more than 528.3 units of resources in February 2012? Q2:: What is the percentage of high consumption customers in February 2012?

Fused question rating: good.

---------------------------------------------------

Question 1507:

SQL:

SELECT DISTINCT T1.Time FROM transactions_1k AS T1 INNER JOIN gasstations AS T2 ON T1.GasStationID = T2.GasStationID WHERE T2.ChainID = 11

Supplied question: Please list the disparate time of the transactions taken place in the gas stations from chain no. 11.

Base generated questions: Q1:: At what distinct times did transactions occur at gas stations belonging to chain ID 11? Q2:: When did gas stations in chain 11 have transactions?

Base question rating: good.

Bird only generated questions: Q1:: What are the distinct times when transactions occurred at gas stations belonging to chain ID 11? Q2:: What times did transactions happen at chain 11 gas stations?

Bird only question rating: good.

LLM only generated questions: **Q1::** What are the distinct times at which transactions occurred at gas stations belonging to the chain with ID 11? **Q2::** What times did transactions happen at gas stations of chain 11?

LLM only question rating: good.

Fused generated questions: Q1:: What are the distinct transaction times for transactions that occurred at gas stations belonging to chain number 11? Q2:: What times did transactions occur at chain 11 gas stations?

Fused question rating: good.

---------------------------------------------------

Question 1525:

SQL:

SELECT CAST(SUM(IIF(T2.Currency = 'EUR', 1, 0)) AS FLOAT) * 100 / COUNT(T1.CustomerID) FROM transactions_1k AS T1 INNER JOIN customers AS T2 ON T1.CustomerID = T2.CustomerID WHERE T1.Date = '2012-08-25'

Supplied question: What is the percentage of the customers who used EUR in 2012/8/25?

Base generated questions: Q1:: What is the percentage of transactions made on August 25, 2012, by customers who have their currency set to EUR, compared to the total number of customers who made transactions on that same date? Q2:: What percent of August 25, 2012 transactions were made by customers using euros?

Base question rating: good+. Reason: suppliedf question does not match the SQL.

Bird only generated questions: Q1:: What is the percentage of transactions made on August 25, 2012, by customers who have their currency set to EUR, relative to the total number of unique customers who made transactions on that date? Q2:: What percentage of August 25, 2012, transactions were by Euro-using customers?

Bird only question rating: good+. Reason: suppliedf question does not match the SQL.

LLM only generated questions: Q1:: What is the percentage of transactions made on August 25, 2012, by customers who have their currency set to EUR, relative to the total number of unique customers who made transactions on that date? Q2:: What percentage of August 25, 2012 transactions were by Euro-using customers?

LLM only question rating: good+. Reason: suppliedf question does not match the SQL.

Fused generated questions: Q1:: What is the percentage of transactions made in euros (EUR) by all customers on the date August 25, 2012? Q2:: What percentage of August 25, 2012, transactions were in euros?

Fused question rating: good+. Reason: suppliedf question does not match the SQL.

---------------------------------------------------

Question 1312:

SQL:

SELECT T2.major_name FROM member AS T1 INNER JOIN major AS T2 ON T1.link_to_major = T2.major_id WHERE T1.first_name = 'Angela' AND T1.last_name = 'Sanders'

Supplied question: What's Angela Sanders's major?

Base generated questions: Q1:: What is the major name associated with the member named Angela Sanders? Q2:: What is Angela Sanders' major?

Base question rating: good.

Bird only generated questions: Q1:: What is the major name associated with a member named Angela Sanders in the database? Q2:: What is Angela Sanders' major?

Bird only question rating: good.

LLM only generated questions: Q1:: What is the major name associated with the member named Angela Sanders? Q2:: What major is Angela Sanders studying?

LLM only question rating: good.

Fused generated questions: Q1:: What is the academic major of the club member Angela Sanders? Q2:: What major is Angela Sanders studying?

Fused question rating: good.

---------------------------------------------------

Question 1338:

SQL:

SELECT T3.approved FROM event AS T1 INNER JOIN budget AS T2 ON T1.event_id = T2.link_to_event INNER JOIN expense AS T3 ON T2.budget_id = T3.link_to_budget WHERE T1.event_name = 'October Meeting' AND T1.event_date LIKE '2019-10-08%'

Supplied question: Was each expense in October Meeting on October 8, 2019 approved?

Base generated questions: Q1:: Was the expense for the 'October Meeting' held on October 8th, 2019 approved? Q2:: Is the October 8th, 2019 'October Meeting' expense approved?

Base question rating: good+. Reason: supplied question doesn't quite match the SQL.

Bird only generated questions: Q1:: Which expenses have been approved for the event named 'October Meeting' that took place on October 8th, 2019? Q2:: What approved expenses are there for the 'October Meeting' on 2019-10-08?

Bird only question rating: good.

LLM only generated questions: Q1:: Was the expense for the 'October Meeting' that occurred on October 8th, 2019 approved? Q2:: Is the October 8th, 2019 'October Meeting' expense approved?

LLM only question rating: good+. Reason: supplied question doesn't quite match the SQL.

Fused generated questions: Q1:: Was the expense for the "October Meeting" that occurred on October 8th, 2019 approved? Q2:: Is the October 8th, 2019 "October Meeting" expense approved?

Fused question rating: good+. Reason: supplied question doesn't quite match the SQL.

---------------------------------------------------

Question 1351:

SQL:

SELECT T2.major_name FROM member AS T1 INNER JOIN major AS T2 ON T1.link_to_major = T2.major_id WHERE T1.first_name = 'Brent' AND T1.last_name = 'Thomason'

Supplied question: What was Brent Thomason's major?

Base generated questions: Q1:: What is the major name associated with the member named Brent Thomason? Q2:: What is Brent Thomason's major?

Base question rating: good.

Bird only generated questions: Q1:: What is the major name associated with the member named Brent Thomason? Q2:: What is Brent Thomason's major?

Bird only question rating: good.

LLM only generated questions: Q1:: What is the name of the major associated with a member named Brent Thomason? Q2:: What major is Brent Thomason studying?

LLM only question rating: good.

Fused generated questions: Q1:: What is the academic major of the club member named Brent Thomason? Q2:: What major is Brent Thomason studying?

Fused question rating: good.

---------------------------------------------------

Question 1362:

SQL:

SELECT COUNT(city) FROM zip_code WHERE county = 'Orange County' AND state = 'Virginia'

Supplied question: How many cities are there in Orange County, Virginia?

Base generated questions: Q1:: How many distinct cities are there in Orange County, Virginia, according to the zip_code database? Q2:: What is the number of cities in Orange County, VA?

Base question rating: good.

Bird only generated questions: Q1:: How many cities are located in Orange County, Virginia according to the zip_code database? Q2:: What is the number of cities in Orange County, VA?

Bird only question rating: good.

LLM only generated questions: Q1:: How many cities are there in Orange County, Virginia, according to the zip_code table? Q2:: What is the city count in Orange County, VA?

LLM only question rating: good.

Fused generated questions: Q1:: How many distinct cities are in Orange County, Virginia, according to the ZIP code data? Q2:: How many cities are in Orange County, VA?

Fused question rating: good.

---------------------------------------------------

Question 1380:

SQL:

SELECT SUM(spent) FROM budget WHERE category = 'Food'

Supplied question: What is the total amount of money spent for food?

Base generated questions: Q1:: How much has been spent in total on the 'Food' category according to the budget records? Q2:: What is the total expenditure on food?

Base question rating: good.

Bird only generated questions: Q1:: What is the total amount of money spent on the 'Food' category according to the records in the budget table? Q2:: How much was spent on food?

Bird only question rating: good.

LLM only generated questions: Q1:: What is the total amount of money spent on the 'Food' category according to the budget records? Q2:: How much was spent on food?

LLM only question rating: good.

Fused generated questions: Q1:: What is the total amount of money spent on the category 'Food' according to the budget records? Q2:: How much was spent on food?

Fused question rating: good.

---------------------------------------------------

Question 1394:

SQL:

SELECT COUNT(T2.member_id) FROM major AS T1 INNER JOIN member AS T2 ON T1.major_id = T2.link_to_major WHERE T1.major_name = 'Physics Teaching'

Supplied question: How many members of the Student_Club have major in 'Physics Teaching'?

Base generated questions: Q1:: How many members are linked to the major with the name 'Physics Teaching'? Q2:: How many students are enrolled in the Physics Teaching major?

Base question rating: good.

Bird only generated questions: Q1:: How many members are associated with the 'Physics Teaching' major according to the database? Q2:: How many members are studying to become physics teachers?

Bird only question rating: good.

LLM only generated questions: Q1:: How many members are associated with the major named 'Physics Teaching'? Q2:: What is the count of members majoring in 'Physics Teaching'?

LLM only question rating: good.

Fused generated questions: Q1:: How many members are associated with the 'Physics Teaching' major according to the database? Q2:: What is the count of 'Physics Teaching' major members?

Fused question rating: good.

---------------------------------------------------

Question 1405:

SQL:

SELECT T2.category, SUM(T2.amount) FROM event AS T1 JOIN budget AS T2 ON T1.event_id = T2.link_to_event WHERE T1.event_name = 'April Speaker' GROUP BY T2.category ORDER BY SUM(T2.amount) ASC

Supplied question: Calculate the amount budgeted for 'April Speaker' event. List all the budgeted categories for said event in an ascending order based on their amount budgeted.

Base generated questions: Q1:: What are the total amounts spent in each category for the event named 'April Speaker', listed in ascending order of amount spent? Q2:: How much was spent per category for 'April Speaker', from least to most?

Base question rating: good.

Bird only generated questions: Q1:: What are the total amounts spent in each category for the event named 'April Speaker', listed in ascending order of expenditure? Q2:: What is the cost breakdown by category for the 'April Speaker' event, from least to most expensive?

Bird only question rating: good.

LLM only generated questions: Q1:: Which categories of the budget are associated with the 'April Speaker' event and what is the total amount spent in each category, listed in ascending order of expenditure? Q2:: What are the spending totals per category for the 'April Speaker' event, from lowest to highest?

LLM only question rating: good.

Fused generated questions: Q1:: What are the total budgeted amounts for each category, listed in ascending order of amount, for the 'April Speaker' event? Q2:: What is the smallest to largest budget allocation by category for the 'April Speaker' event?

Fused question rating: good.

---------------------------------------------------

Question 1427:

SQL:

SELECT DISTINCT T2.category, T1.type FROM event AS T1 INNER JOIN budget AS T2 ON T1.event_id = T2.link_to_event WHERE T1.location = 'MU 215' AND T2.spent = 0 AND T1.type = 'Guest Speaker'

Supplied question: What are the budget category of the events located at MU 215 and a guest speaker type with a 0 budget spent?

Base generated questions: Q1:: What are the distinct categories of budget items linked to events of the type 'Guest Speaker' that have occurred at location 'MU 215' and have a recorded spent amount of 0? Q2:: Which unpaid budget categories are associated with 'Guest Speaker' events at 'MU 215'?

Base question rating: good.

Bird only generated questions: Q1:: What are the distinct categories of budget items associated with events of type 'Guest Speaker' held at location 'MU 215' that have not had any money spent? Q2:: What budget categories are linked to unpaid 'Guest Speaker' events at 'MU 215'?

Bird only question rating: good.

LLM only generated questions: Q1:: What are the distinct categories of budgets linked to events of type 'Guest Speaker' held at location 'MU 215' which have had no expenditure? Q2:: What budget categories are there for unspent 'Guest Speaker' events at 'MU 215'?

LLM only question rating: good.

Fused generated questions: Q1:: What are the distinct budget categories associated with 'Guest Speaker' events that have not had any money spent (budget spent is $0) and are located at 'MU 215'? Q2:: Which budget categories have unused funds for 'Guest Speaker' events at 'MU 215'?

Fused question rating: good.

---------------------------------------------------

Question 1149:

SQL:

SELECT CAST(SUM(CASE WHEN Admission = '+' THEN 1 ELSE 0 END) AS REAL) * 100 / SUM(CASE WHEN Admission = '-' THEN 1 ELSE 0 END) FROM Patient WHERE SEX = 'M'

Supplied question: Are there more in-patient or outpatient who were male? What is the deviation in percentage?

Base generated questions: Step 1: What is the percentage of male patients who were admitted to the hospital compared to the number of male patients who were not admitted? Step 2: What is the admission rate for male patients?

Base question rating: good-. Reason: not picking up on admitted vs. outpatient, but still accurate.

Bird only generated questions: Q1:: What is the percentage of male patients (SEX = 'M') with a positive admission status ('+') compared to the total number of male patients with a negative admission status ('-') in the Patient table? Q2:: What percentage more male patients are admitted than not admitted?

Bird only question rating: good-. Reason: not picking up on admitted vs. outpatient, but still accurate.

LLM only generated questions: Q1:: What is the percentage of male patients admitted (marked with '+') relative to those not admitted (marked with '-')? Q2:: What percentage of male patients were admitted compared to those not admitted?

LLM only question rating: good-. Reason: not picking up on admitted vs. outpatient, but still accurate.

Fused generated questions: Q1:: What is the percentage of male patients who were admitted to the hospital compared to those who were followed at the outpatient clinic? Q2:: What is the hospital admission rate for male patients versus their outpatient clinic visits?

Fused question rating: good.

---------------------------------------------------

Question 1157:

SQL:

SELECT DISTINCT T1.ID, T1.SEX, T1.Diagnosis FROM Patient AS T1 INNER JOIN Examination AS T2 ON T1.ID = T2.ID WHERE T2.Thrombosis = 2

Supplied question: For patients with severe degree of thrombosis, list their ID, sex and disease the patient is diagnosed with.

Base generated questions: Q1:: Which patients have a Thrombosis score of 2, and what are their unique IDs, genders, and diagnoses? Q2:: Who are the patients with a Thrombosis score of 2?

Base question rating: good-. Reason: not picking up on thrombrosis score of 2 meaning severe, but still accurate.

Bird only generated questions: Q1:: Which distinct patients have a diagnosis recorded and have a Thrombosis value of 2 in their examination results? Q2:: Who are the unique patients diagnosed with a Thrombosis level of 2?

Bird only question rating: good-. Reason: not picking up on thrombrosis score of 2 meaning severe, but still accurate.

LLM only generated questions: Q1:: Which patients have a Thrombosis score of 2, and what are their unique IDs, genders, and diagnoses? Q2:: Who are the patients with a Thrombosis score of 2, and what are their genders and diagnoses?

LLM only question rating: good-. Reason: not picking up on thrombrosis score of 2 meaning severe, but still accurate.

Fused generated questions: Q1:: Which patients have a severe degree of thrombosis, as indicated by a thrombosis score of 2, and what are their unique identification numbers, genders, and diagnosed diseases? Q2:: List the ID, sex, and diagnosis of patients with severe thrombosis.

Fused question rating: good.

---------------------------------------------------

Question 1171:

SQL:

SELECT COUNT(T1.ID) FROM Patient AS T1 INNER JOIN Examination AS T2 ON T1.ID = T2.ID WHERE STRFTIME('%Y', T2.`Examination Date`) BETWEEN '1990' AND '1993' AND STRFTIME('%Y', T2.`Examination Date`) - STRFTIME('%Y', T1.Birthday) < 18

Supplied question: How many underage patients were examined during the course of the three-year period from 1990 to 1993?

Base generated questions: Q1:: How many patients were under the age of 18 when they had an examination between the years 1990 and 1993? Q2:: How many minors had examinations from 1990 to 1993?

Base question rating: good.

Bird only generated questions: Q1:: How many patients were under the age of 18 at the time of their examination between the years 1990 and 1993? Q2:: How many minors were examined from 1990 to 1993?

Bird only question rating: good.

LLM only generated questions: Q1:: How many patients were under the age of 18 at the time of their examination that took place between the years 1990 and 1993? Q2:: How many pediatric examinations occurred between 1990 and 1993?

LLM only question rating: good.

Fused generated questions: Q1:: How many patients who were under the age of 18 at the time of their examination had an examination between the years 1990 and 1993? Q2:: How many minors were examined from 1990 to 1993?

Fused question rating: good.

---------------------------------------------------

Question 1192:

SQL:

SELECT DISTINCT T1.ID FROM Patient AS T1 INNER JOIN Laboratory AS T2 ON T1.ID = T2.ID WHERE T1.Admission = '-' AND T2.`T-BIL` < 2.0 AND T2.Date LIKE '1991-10-%'

Supplied question: List all patients who were followed up at the outpatient clinic who underwent a laboratory test in October 1991 and had a total blood bilirubin level within the normal range.

Base generated questions: Q1:: Which patient IDs correspond to individuals who were not admitted, had a total bilirubin level below 2.0, and had laboratory tests conducted in October 1991? Q2:: Who had normal bilirubin levels and lab tests in October 1991 without being admitted?

Base question rating: good-. Reason: not picking up on field meaning details.

Bird only generated questions: Q1:: Which patient IDs correspond to individuals who were not admitted, had a total bilirubin level below 2.0, and underwent laboratory testing in October 1991? Q2:: Who were the outpatients with normal bilirubin levels tested in October 1991?

Bird only question rating: good.

LLM only generated questions: Q1:: Which patient IDs correspond to individuals who were not admitted to the hospital and had a total bilirubin level of less than 2.0 mg/dL during any lab test conducted in October 1991? Q2:: What are the IDs of patients with normal bilirubin levels in October 1991 who were not admitted?

LLM only question rating: good-. Reason: not picking up on field menaing details.

Fused generated questions: Q1:: Which patients were followed at the outpatient clinic and had total bilirubin levels below the normal range in October 1991? Q2:: Who had healthy liver function tests at the outpatient clinic in October 1991?

Fused question rating: good.

---------------------------------------------------

Question 1209:

SQL:

SELECT DISTINCT T1.Diagnosis FROM Patient AS T1 INNER JOIN Laboratory AS T2 ON T1.ID = T2.ID WHERE T2.GPT > 60 ORDER BY T1.Birthday ASC

Supplied question: Please provide the diagnosis of patients with ALT glutamic pylvic transaminase beyond the normal range by ascending order of their date of birth.

Base generated questions: Q1:: What are the distinct diagnoses for patients who have a GPT level greater than 60, sorted by their birthday in ascending order? Q2:: Which unique health conditions are found in patients with elevated GPT levels, listed from youngest to oldest?

Base question rating: good.

Bird only generated questions: Q1:: Which distinct diagnoses are associated with patients who have a GPT level greater than 60, when the patients are sorted by their birth dates in ascending order? Q2:: What are the unique diagnoses for patients with high GPT levels, listed from oldest to youngest?

Bird only question rating: good.

LLM only generated questions: Q1:: What are the distinct diagnoses for patients who have a GPT level greater than 60, sorted by their birthdays in ascending order? Q2:: What distinct conditions have been diagnosed for patients with elevated GPT levels, listed from youngest to oldest?

LLM only question rating: good.

Fused generated questions: Q1:: Which distinct diseases have been diagnosed in patients who have an elevated GPT level above 60, sorted by the patient's age from the youngest to the oldest? Q2:: What diseases are diagnosed in younger to older patients with high liver enzyme levels?

Fused question rating: good.

---------------------------------------------------

Question 1232:

SQL:

SELECT DISTINCT T1.ID, T1.SEX , STRFTIME('%Y', CURRENT_TIMESTAMP) - STRFTIME('%Y', T1.Birthday) FROM Patient AS T1 INNER JOIN Laboratory AS T2 ON T1.ID = T2.ID WHERE T2.GLU >= 180 AND T2.`T-CHO` < 250

Supplied question: Provide ID, sex and age of patient who has blood glucose (GLU) not within normal range but with total cholesterol(T-CHO) within normal range.

Base generated questions: Step 1: Which patients have a glucose level of 180 or higher and a total cholesterol level of less than 250, and what are their IDs, sexes, and current ages? Step 2: Who are the patients with high glucose but not high cholesterol, and what are their IDs, genders, and ages?

Base question rating: good.

Bird only generated questions: Q1:: Which patients have a recorded glucose level of 180 or higher and a total cholesterol level below 250, and what are their IDs, sexes, and current ages? Q2:: Who are the patients with high glucose and acceptable cholesterol levels, and how old are they?

Bird only question rating: good.

LLM only generated questions: Q1:: Which patients have a glucose level of 180 or higher and a total cholesterol level of less than 250, and what are their IDs, sexes, and ages based on the current year? Q2:: Who are the patients with high glucose but not high cholesterol, and what are their ID, gender, and age?

LLM only question rating: good.

Fused generated questions: Q1:: Which patients have a blood glucose level of 180 or higher and a total cholesterol level below 250, and what are their genders and ages? Q2:: Who are the patients with high glucose and acceptable cholesterol levels, along with their genders and ages?

Fused question rating: good.

---------------------------------------------------

Question 1243:

SQL:

SELECT CAST(SUM(CASE WHEN T2.PT >= 14 AND T1.SEX = 'F' THEN 1 ELSE 0 END) AS REAL) * 100 / COUNT(CASE WHEN T2.PT >= 14 THEN 1 ELSE 0 END) FROM Patient AS T1 INNER JOIN Laboratory AS T2 ON T1.ID = T2.ID WHERE STRFTIME('%Y', CURRENT_TIMESTAMP) - STRFTIME('%Y', T1.Birthday) > 55

Supplied question: For all patients who are older than 55 years old, what is the percentage of female who has abnormal prothrombin time (PT)?

Base generated questions: Q1:: What is the percentage of female patients over the age of 55 who have a PT value of 14 or greater, out of all patients over the age of 55 with a PT value of 14 or greater? Q2:: What percentage of older females (over 55) have high PT levels?

Base question rating: good.

Bird only generated questions: Step 1: What percentage of female patients over the age of 55 have a PT (Prothrombin Time) value of 14 or higher? Step 2: What is the percentage of women older than 55 with a high PT level?

Bird only question rating: good.

LLM only generated questions: Q1:: What is the percentage of female patients over the age of 55 who have a PT (Prothrombin

Time) value of 14 or greater out of all patients over the age of 55 with a recorded PT value of 14 or greater? Q2:: What percentage of women over 55 have a high PT level?

LLM only question rating: good.

Fused generated questions: Q1:: What is the percentage of female patients over the age of 55 who have an abnormal prothrombin time (PT) of 14 or higher, out of all patients over the age of 55 with an abnormal PT? Q2:: What percentage of women over 55 have a high blood clotting time?

Fused question rating: good.

---------------------------------------------------

Question 1256:

SQL:

SELECT COUNT(T1.ID) FROM Patient AS T1 INNER JOIN Laboratory AS T2 ON T1.ID = T2.ID WHERE (T2.CRP = '+' ) AND T1.Description IS NULL

Supplied question: How many patients with a abnormal C-reactive protein don't have their data recorded?

Base generated questions: Q1:: How many patients have a positive CRP test result recorded in the Laboratory table and have a NULL description in the Patient table? Q2:: How many patients with positive CRP lack a description?

Base question rating: good.

Bird only generated questions: Q1:: How many patients have a positive C-reactive protein (CRP) test result recorded in the Laboratory table and do not have a description entered in the Patient table? Q2:: How many patients with positive CRP lack a description?

Bird only question rating: good.

LLM only generated questions: Q1:: How many patients have a positive C-reactive protein (CRP) test result in the Laboratory table and have a null description in the Patient table? Q2:: How many patients with positive CRP lack a description?

LLM only question rating: good.

Fused generated questions: Q1:: How many patients have a positive C-reactive protein test result indicating inflammation but do not have a recorded first date of data entry in the system? Q2:: How many patients with inflammation have no recorded first data entry?

Fused question rating: good.

---------------------------------------------------

Question 1281:

SQL:

SELECT T1.Birthday FROM Patient AS T1 INNER JOIN Laboratory AS T2 ON T1.ID = T2.ID WHERE T2.GOT >= 60 ORDER BY T1.Birthday DESC LIMIT 1

Supplied question: Among the patients who have an abnormal level of glutamic oxaloacetic transaminase, when was the youngest of them born?

Base generated questions: Q1:: What is the most recent birthday of a patient who has a Glutamic Oxaloacetic Transaminase (GOT) level of 60 or higher as recorded in the laboratory results? Q2:: Who is the youngest patient with a GOT level at or above 60?

Base question rating: good-. Reason: Not picking up on field meanings.

Bird only generated questions: Q1:: What is the most recent birthday among patients who have a Glutamic-Oxaloacetic Transaminase (GOT) level of 60 or higher according to the laboratory results? Q2:: Who is the youngest patient with a GOT level of at least 60?

Bird only question rating: good-. Reason: Not picking up on field meanings.

LLM only generated questions: **Q1::** What is the most recent birthday of a patient who has a Glutamate Oxaloacetate Transaminase (GOT) level of 60 or higher according to laboratory records? **Q2::** What is the latest birthday for a patient with a GOT level at least 60?

LLM only question rating: good-. Reason: Not picking up on field meanings.

Fused generated questions: Q1:: What is the most recent birthday of a patient who has an AST level of 60 or higher? Q2:: Who is the youngest patient with elevated AST levels?

Fused question rating: good.

---------------------------------------------------

Question 1031:

SQL:

SELECT DISTINCT DATETIME() - T2.birthday age FROM Player_Attributes AS t1 INNER JOIN Player AS t2 ON t1.player_api_id = t2.player_api_id WHERE STRFTIME('%Y',t1.`date`) >= '2013' AND STRFTIME('%Y',t1.`date`) <= '2015' AND t1.sprint_speed >= 97

Supplied question: At present, calculate for the player's age who have a sprint speed of no less than 97 between 2013 to 2015.

Base generated questions: Q1:: What are the distinct ages of players who had a sprint speed of 97 or higher recorded between the years 2013 and 2015? Q2:: What ages did the fastest players have during 2013-2015?

Base question rating: good.

Bird only generated questions: Q1:: Which players had a sprint speed of 97 or higher between the years 2013 and 2015, and what were their ages at the time of the recorded sprint speed? Q2:: Who were the fastest players aged 97 or above from 2013 to 2015?

Bird only question rating: good-. Reason: getting ID, not age.

LLM only generated questions: Q1:: What are the distinct ages of players who had a sprint speed of 97 or higher recorded in their player attributes between the years 2013 and 2015? Q2:: How old were the fastest players between 2013 and 2015?

LLM only question rating: good.

Fused generated questions: Q1:: What are the ages of players who have a sprint speed rating of 97 or higher according to the attributes recorded between the years 2013 and 2015? Q2:: How old are the fastest players, with speeds of at least 97, based on data from 2013 to 2015?

Fused question rating: good.

---------------------------------------------------

Question 1040:

SQL:

SELECT t1.player_name FROM Player AS t1 INNER JOIN Player_Attributes AS t2 ON t1.player_api_id = t2.player_api_id WHERE t1.height > 180 GROUP BY t1.id ORDER BY

CAST(SUM(t2.heading_accuracy) AS REAL) / COUNT(t2.`player_fifa_api_id`) DESC LIMIT 10

Supplied question: List the top 10 players' names whose heights are above 180 in descending order of average heading accuracy.

Base generated questions: Q1:: Which ten players taller than 180 cm have the highest average heading accuracy in the Player_Attributes table? Q2:: Who are the top ten tallest players with the best average heading skills?

Base question rating: good-. Reason: not picking up on the order by.

Bird only generated questions: Q1:: Which ten players taller than 180 cm have the highest average heading accuracy in the Player_Attributes table? Q2:: Who are the top ten tallest players with the best average heading skill?

Bird only question rating: good-. Reason: not picking up on the order by.

LLM only generated questions: Q1:: Which ten players taller than 180 cm have the highest average heading accuracy in the Player_Attributes table? Q2:: Who are the top ten tallest players with the best average heading accuracy?

LLM only question rating: good-. Reason: not picking up on the order by.

Fused generated questions: Q1:: Which 10 players taller than 180 cm have the highest average heading accuracy score in the database, based on their attributes? Q2:: Who are the top 10 tallest players with the best average heading accuracy?

Fused question rating: good-. Reason: not picking up on the order by.

---------------------------------------------------

Question 1068:

SQL:

SELECT CAST(SUM(t2.overall_rating) AS REAL) / COUNT(t2.id) FROM Player AS t1 INNER JOIN Player_Attributes AS t2 ON t1.player_api_id = t2.player_api_id WHERE t1.height > 170 AND STRFTIME('%Y',t2.`date`) >= '2010' AND STRFTIME('%Y',t2.`date`) <= '2015'

Supplied question: From 2010 to 2015, what was the average overall rating of players who are higher than 170?

Base generated questions: Q1:: What is the average overall rating of players who are taller than 170 cm, based on the player attributes recorded between the years 2010 and 2015? Q2:: What's the average rating for players over 170 cm tall from 2010 to 2015?

Base question rating: good.

Bird only generated questions: Q1:: What is the average overall rating of players who are taller than 170 cm, based on their attributes recorded between the years 2010 and 2015? Q2:: What's the average rating for players over 170 cm tall from 2010 to 2015?

Bird only question rating: good.

LLM only generated questions: **Q1::** What is the average overall rating of players who are taller than 170 cm, based on their attributes recorded between the years 2010 and 2015? **Q2::** What is the average rating for players over 170 cm tall from 2010 to 2015?

LLM only question rating: good.

Fused generated questions: Q1:: What is the average overall rating for players taller than 170 cm based on the recorded attributes between the years 2010 and 2015? Q2:: What is the average FIFA strength rating for players over 170 cm tall from 2010 to 2015?

Fused question rating: good.

---------------------------------------------------

Question 1088:

SQL:

SELECT DISTINCT t1.player_name FROM Player AS t1 INNER JOIN Player_Attributes AS t2 ON t1.player_api_id = t2.player_api_id WHERE t2.volleys > 70 AND t2.dribbling > 70

Supplied question: Please list the names of the players whose volley score and dribbling score are over 70.

Base generated questions: Q1:: Which players have a volleys attribute greater than 70 and a dribbling attribute also greater than 70, without listing any player more than once? Q2:: Who are the players with both volleys and dribbling over 70?

Base question rating: good.

Bird only generated questions: Q1:: Which players have a volleys attribute greater than 70 and a dribbling attribute greater than 70? Q2:: Who are the players with volleys and dribbling skills over 70?

Bird only question rating: good.

LLM only generated questions: Q1:: Which players have both volleys and dribbling attributes greater than 70? Q2:: Who are the players with volleys and dribbling skills above 70?

LLM only question rating: good.

Fused generated questions: Q1:: Which players have a volley score and a dribbling score both greater than 70 according to the FIFA calculations? Q2:: Who are the players with volley and dribbling skills above 70?

Fused question rating: good.

---------------------------------------------------

Question 1102:

SQL:

SELECT t1.player_name FROM Player AS t1 INNER JOIN Player_Attributes AS t2 ON t1.player_api_id = t2.player_api_id WHERE SUBSTR(t2.`date`, 1, 10) = '2016-06-23' AND t2.overall_rating = 77 ORDER BY t1.birthday ASC LIMIT 1

Supplied question: For the players who had a 77 points overall rating on 2016/6/23, who was the oldest? Give the name of the player.

Base generated questions: Q1:: Who is the earliest born player with an overall rating of 77 on June 23, 2016, according to the Player and Player_Attributes tables? Q2:: Who was the oldest player with a rating of 77 on June 23, 2016?

Base question rating: good.

Bird only generated questions: Q1:: Who is the earliest born player with an overall rating of 77 on the date of June 23, 2016, according to the Player and Player_Attributes tables? Q2:: Who was the oldest player with a 77 rating on June 23, 2016?

Bird only question rating: good.

LLM only generated questions: Q1:: Who is the earliest born player with an overall rating of 77 on June 23, 2016, according to the Player and Player_Attributes tables? Q2:: Who is the oldest player with a rating of 77 on June 23, 2016?

LLM only question rating: good.

Fused generated questions: Q1:: Who is the oldest player with an overall rating of 77 according to the player attributes recorded on June 23, 2016? Q2:: Which oldest player had a FIFA rating of 77 on June 23, 2016?

Fused question rating: good.

---------------------------------------------------

Question 1114:

SQL:

SELECT CAST(SUM(t2.overall_rating) AS REAL) / COUNT(t2.id) FROM Player AS t1 INNER JOIN Player_Attributes AS t2 ON t1.player_fifa_api_id = t2.player_fifa_api_id WHERE t1.player_name = 'Marko Arnautovic' AND SUBSTR(t2.`date`, 1, 10) BETWEEN '2007-02-22' AND '2016-04-21'

Supplied question: What was the average overall rating for Marko Arnautovic from 2007/2/22 to 2016/4/21?

Base generated questions: Q1:: What is the average overall rating of the player named 'Marko Arnautovic' in the Player_Attributes table, based on records from February 22, 2007, to April 21, 2016? Q2:: What is Marko Arnautovic's average rating between 2007 and 2016?

Base question rating: good.

Bird only generated questions: Q1:: What is the average overall rating of the player named 'Marko Arnautovic' according to the Player_Attributes table for the period between February 22, 2007, and April 21, 2016? Q2:: What is Marko Arnautovic's average rating from 2007 to 2016?

Bird only question rating: good.

LLM only generated questions: Q1:: What is the average overall rating of the player named 'Marko Arnautovic' according to the Player_Attributes table, for the records that fall between the dates '2007-02-22' and '2016-04-21'? Q2:: What is Marko Arnautovic's average rating from 2007 to 2016?

LLM only question rating: good.

Fused generated questions: Q1:: What is the average overall rating of the player named 'Marko Arnautovic' according to the FIFA database for the period between February 22, 2007, and April 21, 2016? Q2:: What is Marko Arnautovic's average FIFA rating from 2007 to 2016?

Fused question rating: good.

---------------------------------------------------

Question 1133:

SQL:

SELECT COUNT(id) FROM Player WHERE birthday > '1990' AND player_name LIKE 'Aaron%'

Supplied question: How many football players born after the 1990s have the first name "Aaron"?

Base generated questions: Q1:: How many players named Aaron who were born after the year 1990 are recorded in the Player database? Q2:: How many post-1990 Aarons are in the Player database?

Base question rating: good.

Bird only generated questions: Q1:: How many players in the Player table were born after the year 1990 and have a name that starts with "Aaron"? Q2:: How many "Aaron"s born post-1990 are listed as players?

Bird only question rating: good.

LLM only generated questions: Q1:: How many players named Aaron, or with a name starting with "Aaron", were born after the year 1990 according to the database? Q2:: How many Aarons are born post-1990?

LLM only question rating: good.

Fused generated questions: Q1:: How many players in the database were born after the year 1990 and have a name that starts with "Aaron"? Q2:: How many post-1990 players are named Aaron?

Fused question rating: good.

---------------------------------------------------

Question 1144:

SQL:

SELECT id, finishing, curve FROM Player_Attributes WHERE player_api_id = ( SELECT player_api_id FROM Player ORDER BY weight DESC LIMIT 1 ) LIMIT 1

Supplied question: Please state the finishing rate and curve score of the player who has the heaviest weight.

Base generated questions: Q1:: What are the id, finishing, and curve attributes of the player who has the highest weight in the Player_Attributes table? Q2:: What are the key stats for the heaviest player?

Base question rating: good.

Bird only generated questions: Q1:: What are the id, finishing, and curve attributes of the player with the highest weight in the Player_Attributes table? Q2:: What skills does the heaviest player have?

Bird only question rating: good.

LLM only generated questions: Q1:: What are the id, finishing, and curve attributes of the player who has the highest weight according to the Player_Attributes table? Q2:: What are the key attributes of the heaviest player?

LLM only question rating: good.

Fused generated questions: Q1:: What are the id, finishing, and curve attributes for the player with the highest weight in the Player table? Q2:: What finishing and curve scores does the heaviest player have?

Fused question rating: good.

---------------------------------------------------

Question 847:

SQL:

SELECT T2.surname FROM qualifying AS T1 INNER JOIN drivers AS T2 ON T2.driverId = T1.driverId WHERE T1.raceId = 19 ORDER BY T1.q2 ASC LIMIT 1

Supplied question: What is the surname of the driver with the best lap time in race number 19 in the second qualifying period?

Base generated questions: Q1:: Which driver's surname corresponds to the driver who achieved the fastest time in the second qualifying session (Q2) for the race with the raceId of 19? Q2:: Who had the fastest Q2 time in race 19?

Base question rating: good.

Bird only generated questions: Q1:: Which driver's surname is associated with the fastest qualifying time in Q2 for race number 19? Q2:: Who had the fastest Q2 time in race 19?

Bird only question rating: good.

LLM only generated questions: Q1:: Who is the driver with the surname that qualified fastest in the second qualifying session for race number 19? Q2:: Which driver had the best Q2 time in the 19th race?

LLM only question rating: good.

Fused generated questions: Q1:: Which driver's surname corresponds to the driver who had the fastest lap time in the second qualifying session (Q2) for the race with the identification number 19? Q2:: Who had the fastest Q2 lap in race 19?

Fused question rating: good.

---------------------------------------------------

Question 862:

SQL:

SELECT COUNT(T3.driverId) FROM races AS T1 INNER JOIN results AS T2 ON T2.raceId = T1.raceId INNER JOIN drivers AS T3 ON T3.driverId = T2.driverId WHERE T1.year = 2007 AND T1.name = 'Bahrain Grand Prix' AND T2.time IS NULL

Supplied question: For the Bahrain Grand Prix in 2007, how many drivers not finished the game?

Base generated questions: Q1:: How many drivers who participated in the 2007 Bahrain Grand Prix did not finish the race? Q2:: How many drivers did not finish the 2007 Bahrain Grand Prix?

Base question rating: good+. Reason: the supplied question is not proper English.

Bird only generated questions: Step 1: How many drivers who participated in the 2007 Bahrain Grand Prix did not finish the race (as indicated by having a NULL time in the results)? Step 2: How many drivers did not finish the 2007 Bahrain Grand Prix?

Bird only question rating: good+. Reason: the supplied question is not proper English.

LLM only generated questions: Q1:: How many drivers who participated in the 2007 Bahrain Grand Prix did not finish the race, as indicated by having no recorded finish time in the results? Q2:: How many drivers did not finish the 2007 Bahrain Grand Prix?

LLM only question rating: good+. Reason: the supplied question is not proper English.

Fused generated questions: Q1:: How many drivers who participated in the 2007 Bahrain Grand Prix did not finish the race? Q2:: How many drivers did not finish the 2007 Bahrain Grand Prix?

Fused question rating: good+. Reason: the supplied question is not proper English.

---------------------------------------------------

Question 875:

SQL:

SELECT T2.url FROM races AS T1 INNER JOIN seasons AS T2 ON T2.year = T1.year WHERE T1.raceId = 901

Supplied question: Show me the season page of year when the race No. 901 took place.

Base generated questions: Q1:: What is the URL of the season associated with the race that has the raceId of 901? Q2:: What's the season's URL for race 901?

Base question rating: good.

Bird only generated questions: Q1:: What is the URL of the season associated with the race that has the raceId of 901? Q2:: What's the season's URL for the race with ID 901?

Bird only question rating: good.

LLM only generated questions: Q1:: What is the URL of the season corresponding to the race with the raceId of 901 in the database? Q2:: What is the season URL for race 901?

LLM only question rating: good.

Fused generated questions: Q1:: What is the website link for the introduction page of the Formula One season in which the race with the raceId 901 occurred? Q2:: What's the intro page link for the F1 season of race 901?

Fused question rating: good. Reason: more accurate but not really needed.

---------------------------------------------------

Question 892:

SQL:

SELECT T3.forename, T3.surname, T2.points FROM races AS T1 INNER JOIN driverStandings AS T2 ON T2.raceId = T1.raceId INNER JOIN drivers AS T3 ON T3.driverId = T2.driverId ORDER BY T2.points DESC LIMIT 1

Supplied question: State the driver with the most points scored. Find his full name with that points.

Base generated questions: Q1:: Which driver has the highest number of points in a specific race, and what are their first and last names? Q2:: Who is the top-scoring driver in a race?

Base question rating: good-. Reason: a race vs. any race.

Bird only generated questions: Q1:: Which driver has the highest number of points in a single race, and what are their forename and surname? Q2:: Who is the top-scoring driver in any race?

Bird only question rating: good.

LLM only generated questions: Q1:: Which driver has the highest number of points, and what are their forename and surname, based on the latest race results in the database? Q2:: Who is the top-scoring driver in the latest race?

LLM only question rating: good. Reason: focusing on latest race.

Fused generated questions: Q1:: Which driver, including their first and last name, scored the highest number of points in a single race, according to the records in the database? Q2:: Who scored the most points in a single race?

Fused question rating: good.

---------------------------------------------------

Question 901:

SQL:

SELECT DISTINCT T2.name, T1.name, T1.location FROM circuits AS T1 INNER JOIN races AS T2 ON T2.circuitID = T1.circuitId WHERE T2.year = 2005 AND STRFTIME('%m', T2.date) = '09'

Supplied question: Name the races along with its circuit name and location for f1 races hosted in September 2005.

Base generated questions: Q1:: What are the names of the circuits and the corresponding race names for races that took place in September 2005, along with the locations of those circuits? Q2:: Which races and circuits occurred in September 2005, and where were they located?

Base question rating: good.

Bird only generated questions: Q1:: What are the names and locations of the circuits and the corresponding names of the races that took place in September 2005? Q2:: Which races and circuits occurred in September 2005?

Bird only question rating: good-. Reason: not saying all the output fields.

LLM only generated questions: Q1:: Which race names and circuit names with their locations are associated with races that took place in September 2005? Q2:: What are the names and locations of circuits that hosted races in September 2005?

LLM only question rating: good.

Fused generated questions: Q1:: What are the full names of the racing circuits and their locations, along with the official names of the races that took place in September 2005? Q2:: Which races and circuits were in the September 2005 schedule?

Fused question rating: good.

---------------------------------------------------

Question 912:

SQL:

SELECT circuitRef FROM circuits WHERE name = 'Marina Bay Street Circuit'

Supplied question: What's the reference name of Marina Bay Street Circuit?

Base generated questions: Q1:: What is the circuit reference code for the racing track known as 'Marina Bay Street Circuit' in the database? Q2:: What's the code for the 'Marina Bay Street Circuit'?

Base question rating: good.

Bird only generated questions: Q1:: What is the circuit reference code for the 'Marina Bay Street Circuit' in the circuits table? Q2:: What is the 'Marina Bay Street Circuit' reference code?

Bird only question rating: good.

LLM only generated questions: Q1:: What is the circuit reference code for the 'Marina Bay Street Circuit' in the circuits database? Q2:: What's the reference code for Marina Bay Street Circuit?

LLM only question rating: good.

Fused generated questions: Q1:: What is the unique reference name for the racing circuit with the full name 'Marina Bay Street Circuit' in the database? Q2:: What is the reference name for 'Marina Bay Street Circuit'?

Fused question rating: good.

---------------------------------------------------

Question 937:

SQL:

SELECT T1.time FROM results AS T1 INNER JOIN races AS T2 on T1.raceId = T2.raceId WHERE T1.rank = 2 AND T2.name = 'Chinese Grand Prix' AND T2.year = 2008

Supplied question: What's the finish time for the driver who ranked second in 2008's AustChineseralian Grand Prix?

Base generated questions: Q1:: What was the time recorded for the driver who finished in second place at the Chinese Grand Prix in the year 2008? Q2:: What was the runner-up's time in the 2008 Chinese Grand Prix?

Base question rating: good+. Reason: bad typo in supplied question.

Bird only generated questions: Q1:: What was the time achieved by the driver who finished in second place at the Chinese Grand Prix in the year 2008? Q2:: What was the runner-up's time in the 2008 Chinese Grand Prix?

Bird only question rating: good+. Reason: bad typo in supplied question.

LLM only generated questions: Q1:: What was the time taken by the driver who finished in second place in the Chinese Grand Prix in the year 2008? Q2:: What was the runner-up's time in the 2008 Chinese Grand Prix?

LLM only question rating: good+. Reason: bad typo in supplied question.

Fused generated questions: Q1:: What was the finish time difference of the driver who ranked second in the Chinese Grand Prix in 2008? Q2:: What was the runner-up's time behind the winner in the 2008 Chinese Grand Prix?

Fused question rating: good+. Reason: bad typo in supplied question.

---------------------------------------------------

Question 951:

SQL:

SELECT COUNT(T1.raceId) FROM constructorStandings AS T1 INNER JOIN constructors AS T2 on T1.constructorId = T2.constructorId WHERE T1.points = 0 AND T2.nationality = 'Japanese' GROUP BY T1.constructorId HAVING COUNT(raceId) = 2

Supplied question: How many Japanese constructors have 0 points in 2 races?

Base generated questions: Q1:: How many Japanese constructors have exactly two race entries where their teams scored zero points? Q2:: Which Japanese constructors have two races with no points scored?

Base question rating: good.

Bird only generated questions: Q1:: How many Japanese constructors have participated exactly in two races without scoring any points? Q2:: Which Japanese constructors have two zero-point races?

Bird only question rating: good.

LLM only generated questions: Q1:: How many Japanese constructors have exactly two race entries where their teams scored zero points? Q2:: Which Japanese constructors had two races with no points scored?

LLM only question rating: good.

Fused generated questions: Q1:: Which Japanese constructors have exactly two race entries where they scored zero points? Q2:: Which Japanese constructors finished two races with no points?

Fused question rating: good.

---------------------------------------------------

Question 963:

SQL:

SELECT COUNT(T1.driverId) FROM drivers AS T1 INNER JOIN lapTimes AS T2 on T1.driverId = T2.driverId WHERE T1.nationality = 'French' AND (CAST(SUBSTR(T2.time, 1, 2) AS INTEGER) * 60 + CAST(SUBSTR(T2.time, 4, 2) AS INTEGER) + CAST(SUBSTR(T2.time, 7, 2) AS REAL) / 1000) < 120

Supplied question: How many French drivers who obtain the laptime less than 02:00.00?

Base generated questions: Q1:: How many French drivers have recorded a lap time of less than two minutes in the lapTimes database?  Q2:: How many French drivers completed a lap in under two minutes?

Base question rating: good+.  Reason: supplied question is ambiguous about count vs. name.

Bird only generated questions: Q1:: How many laps have been completed by French drivers in the lapTimes table where their lap time was less than two minutes?  Q2:: How many sub-two-minute laps were completed by French drivers?

Bird only question rating: good+.  Reason: supplied question is ambiguous about count vs. name.

LLM only generated questions: Q1:: How many French drivers have completed a lap in less than 120 seconds according to the lapTimes database?  Q2:: How many French drivers have lapped under 2 minutes?

LLM only question rating: good+.  Reason: supplied question is ambiguous about count vs. name.

Fused generated questions: Q1:: How many French drivers have completed a lap in less than 120 seconds?  Q2:: How many French drivers finished a lap in under 2 minutes?

Fused question rating: good+.  Reason: supplied question is ambiguous about count vs. name.

---------------------------------------------------

Question 978:

SQL:

SELECT DISTINCT location, lat, lng FROM circuits WHERE country = 'Austria'

Supplied question: How many times the circuits were held in Austria? Please give their location and coordinates.

Base generated questions: Q1:: Which unique locations and their corresponding latitude and longitude coordinates are associated with circuits in Austria?  Q2:: What are the distinct locations and coordinates of Austrian racing circuits?

Base question rating: good+.  Reason: supplied question is ambiguous between count and list.

Bird only generated questions: Q1:: What are the unique locations and their latitude and longitude coordinates for the racing circuits located in Austria?  Q2:: Where are Austria's racing circuits and their coordinates?

Bird only question rating: good+.  Reason: supplied question is ambiguous between count and list.

LLM only generated questions: Q1:: What are the unique locations and their corresponding latitude and longitude coordinates for the racing circuits located in Austria?  Q2:: What are the distinct locations and GPS coordinates of racing tracks in Austria?

LLM only question rating: good+.  Reason: supplied question is ambiguous between count and list.

Fused generated questions: Q1:: What are the unique names of the cities along with their latitude and longitude coordinates where the racing circuits are located in Austria?  Q2:: What are the distinct city names and coordinates of racing circuits in Austria?

Fused question rating: good+.  Reason: supplied question is ambiguous between count and list.

---------------------------------------------------

Question 994:

SQL:

SELECT SUM(T1.points), T2.name, T2.nationality FROM constructorResults AS T1 INNER JOIN constructors AS T2 ON T1.constructorId = T2.constructorId INNER JOIN races AS T3 ON T3.raceid = T1.raceid WHERE T3.name = 'Monaco Grand Prix' AND T3.year BETWEEN 1980 AND 2010 GROUP BY T2.name ORDER BY SUM(T1.points) DESC LIMIT 1

Supplied question: Which constructor scored most points from Monaco Grand Prix between 1980 and 2010? List the score, name and nationality of this team.

Base generated questions: Q1:: Which constructor has the highest total number of points scored at the Monaco Grand Prix between the years 1980 and 2010, and what is the name and nationality of that constructor?  Q2:: Who was the most successful constructor at the Monaco Grand Prix from 1980 to 2010, and where were they from?

Base question rating: good.  Reason: more succinct but does not explicitly ask for the score.

Bird only generated questions: Q1:: Which constructor has scored the highest total points at the Monaco Grand Prix between the years 1980 and 2010, and what is the name and nationality of that constructor?  Q2:: Who was the top-scoring constructor from 1980 to 2010 at the Monaco Grand Prix, and where were they from?

Bird only question rating: good.  Reason: more succinct but does not explicitly ask for the score.

LLM only generated questions: Q1:: Which constructor has scored the highest total points at the Monaco Grand Prix from 1980 to 2010, and what is the name and nationality of that constructor?  Q2:: Who was the most successful constructor at the Monaco Grand Prix between 1980 and 2010, and where were they from?

LLM only question rating: good.  Reason: more succinct but does not explicitly ask for the score.

Fused generated questions: Q1:: Which constructor scored the highest total points at the Monaco Grand Prix from 1980 to 2010, and what is their nationality?  Q2:: Who was the top-scoring constructor at the Monaco Grand Prix between 1980 and 2010, and where were they from?

Fused question rating: good.  Reason: more succinct but does not explicitly ask for the score.

---------------------------------------------------

Question 719:

SQL:

SELECT COUNT(T1.id) FROM superhero AS T1 INNER JOIN hero_power AS T2 ON T1.id = T2.hero_id INNER JOIN superpower AS T3 ON T2.power_id = T3.id WHERE T3.power_name = 'Super Strength' AND T1.height_cm > 200

Supplied question: Among the superheroes with the super power of "Super Strength", how many of them have a height of over 200cm?

Base generated questions: Q1:: How many superheroes who are taller than 200 centimeters possess the power of 'Super Strength' according to the database?  Q2:: How many tall superheroes have super strength?

Base question rating: good.

Bird only generated questions: Q1:: How many superheroes who are taller than 200 centimeters possess the 'Super Strength' power according to the database? Q2:: How many tall superheroes have super strength?

Bird only question rating: good.

LLM only generated questions: Q1:: How many superheroes are taller than 200 centimeters and possess the superpower of 'Super Strength'? Q2:: How many tall superheroes have super strength?

LLM only question rating: good.

Fused generated questions: Q1:: How many superheroes with a height greater than 200 centimeters possess the superpower 'Super Strength'? Q2:: How many tall superheroes (over 200 cm) have super strength?

Fused question rating: good.

---------------------------------------------------

Question 733:

SQL:

SELECT COUNT(T1.id) FROM superhero AS T1 INNER JOIN publisher AS T2 ON T1.publisher_id = T2.id INNER JOIN colour AS T3 ON T1.eye_colour_id = T3.id WHERE T2.publisher_name = 'Marvel Comics' AND T3.colour = 'Gold'

Supplied question: How many gold-eyed superheroes did Marvel Comics publish?

Base generated questions: Q1:: How many superheroes in the database have a publisher named 'Marvel Comics' and have gold as their eye color? Q2:: How many Marvel superheroes have gold eyes?

Base question rating: good.

Bird only generated questions: Q1:: How many superheroes in the database have a publisher named 'Marvel Comics' and gold eye colour? Q2:: How many Marvel superheroes have gold eyes?

Bird only question rating: good.

LLM only generated questions: Q1:: How many superheroes in the database have a publisher named 'Marvel Comics' and gold eye colour? Q2:: How many Marvel superheroes have gold eyes?

LLM only question rating: good.

Fused generated questions: Q1:: How many superheroes published by 'Marvel Comics' have 'Gold' as their eye color? Q2:: How many Marvel superheroes have gold eyes?

Fused question rating: good.

---------------------------------------------------

Question 744:

SQL:

SELECT SUM(CASE WHEN T2.publisher_name = 'Marvel Comics' THEN 1 ELSE 0 END) - SUM(CASE WHEN T2.publisher_name = 'DC Comics' THEN 1 ELSE 0 END) FROM superhero AS T1 INNER JOIN publisher AS T2 ON T1.publisher_id = T2.id

Supplied question: Between DC and Marvel Comics, which publisher has published more superheroes? Find the difference in the number of superheroes they have published.

Base generated questions: Q1:: What is the difference in the number of superheroes published by Marvel Comics and DC Comics according to the database? Q2:: How many more superheroes does Marvel have compared to DC?

Base question rating: good+. Reason: supplied question asks for name of more prolific publisher, this is not part of the SQL.

Bird only generated questions: Q1:: What is the difference in the number of superheroes published by Marvel Comics and DC Comics according to the database? Q2:: How many more superheroes does Marvel have compared to DC?

Bird only question rating: good+. Reason: supplied question asks for name of more prolific publisher, this is not part of the SQL.

LLM only generated questions: Q1:: What is the difference in the number of superheroes published by Marvel Comics and DC Comics according to the database? Q2:: How many more superheroes does Marvel have compared to DC?

LLM only question rating: good+. Reason: supplied question asks for name of more prolific publisher, this is not part of the SQL.

Fused generated questions: Q1:: How many more superheroes are published by Marvel Comics than by DC Comics according to the database? Q2:: What is the difference in the number of superheroes between Marvel Comics and DC Comics?

Fused question rating: good+. Reason: supplied question asks for name of more prolific publisher, this is not part of the SQL.

---------------------------------------------------

Question 758:

SQL:

SELECT DISTINCT T3.colour FROM superhero AS T1 INNER JOIN race AS T2 ON T1.race_id = T2.id INNER JOIN colour AS T3 ON T1.hair_colour_id = T3.id WHERE T1.height_cm = 185 AND T2.race = 'Human'

Supplied question: Provide the hair colour of the human superhero who is 185 cm tall.

Base generated questions: Q1:: What are the distinct hair colours of human superheroes who are exactly 185 cm tall? Q2:: What hair colours do 185 cm tall human superheroes have?

Base question rating: good+. Reason: question implies singular, query can return zero or more.

Bird only generated questions: Q1:: What are the distinct hair colours of human superheroes who are exactly 185 cm tall? Q2:: What hair colours do 185 cm tall human superheroes have?

Bird only question rating: good+. Reason: question implies singular, query can return zero or more.

LLM only generated questions: Q1:: What are the distinct hair colours of human superheroes who are exactly 185 cm tall? Q2:: What hair colours do 185 cm tall human superheroes have?

LLM only question rating: good+. Reason: question implies singular, query can return zero or more.

Fused generated questions: Q1:: What are the distinct hair colors of human superheroes who are exactly 185 centimeters tall? Q2:: What hair colors do 185 cm tall human superheroes have?

Fused question rating: good+. Reason: question implies singular, query can return zero or more.

---------------------------------------------------

Question 769:

SQL:

SELECT T1.superhero_name FROM superhero AS T1 INNER JOIN hero_attribute AS T2 ON T1.id = T2.hero_id INNER JOIN attribute AS T3 ON T3.id = T2.attribute_id INNER JOIN publisher AS T4 ON T4.id = T1.publisher_id WHERE T4.publisher_name = 'Dark Horse Comics' AND T3.attribute_name = 'Durability' ORDER BY T2.attribute_value DESC LIMIT 1

Supplied question: Which superhero has the most durability published by Dark Horse Comics?

Base generated questions: Q1:: Which superhero from Dark Horse Comics has the highest durability attribute value?  Q2:: Who is the most durable superhero from Dark Horse Comics?

Base question rating: good.  Reason: poor grammer in the supplied question.

Bird only generated questions: Q1:: Which superhero from Dark Horse Comics has the highest durability attribute value?  Q2:: Who is the most durable superhero from Dark Horse Comics?

Bird only question rating: good.  Reason: poor grammer in the supplied question.

LLM only generated questions: Q1:: Which superhero from 'Dark Horse Comics' has the highest 'Durability' attribute value?  Q2:: Who is the most durable superhero from Dark Horse Comics?

LLM only question rating: good.  Reason: poor grammer in the supplied question.

Fused generated questions: Q1:: Which superhero from 'Dark Horse Comics' has the highest 'Durability' attribute value?  Q2:: Who is the most durable superhero from Dark Horse Comics?

Fused question rating: good.  Reason: poor grammer in the supplied question.

---------------------------------------------------

Question 782:

SQL:

SELECT T1.superhero_name FROM superhero AS T1 INNER JOIN colour AS T2 ON T1.eye_colour_id = T2.id AND T1.hair_colour_id = T2.id WHERE T2.colour = 'Black'

Supplied question: List the heroes' names whose eyes and hair colours are both black.

Base generated questions: Q1:: Which superheroes have both black eye color and black hair color according to the database?  Q2:: Who are the superheroes with black eyes and hair?

Base question rating: good.

Bird only generated questions: Q1:: Which superheroes have both black eye color and black hair color according to the database?  Q2:: Who are the superheroes with black eyes and hair?

Bird only question rating: good.

LLM only generated questions: Q1:: Which superheroes have both eye and hair colour matching with the colour 'Black' in the database?  Q2:: Who are the superheroes with black eyes and hair?

LLM only question rating: good.

Fused generated questions: Q1:: Which superheroes have both their eye color and hair color listed as black in the database?  Q2:: Who are the superheroes with black eyes and hair?

Fused question rating: good.

---------------------------------------------------

Question 792:

SQL:

SELECT T3.power_name FROM superhero AS T1 INNER JOIN hero_power AS T2 ON T1.id = T2.hero_id INNER JOIN superpower AS T3 ON T2.power_id = T3.id WHERE T1.superhero_name = 'Abomination'

Supplied question: What is Abomination's superpower?

Base generated questions: Q1:: What are the names of the superpowers associated with the superhero known as 'Abomination' in the database?  Q2:: What superpowers does Abomination have?

Base question rating: good.

Bird only generated questions: Q1:: What are the names of the superpowers associated with the superhero known as 'Abomination' in the database?  Q2:: What superpowers does Abomination have?

Bird only question rating: good.

LLM only generated questions: Q1:: What are the names of the superpowers associated with the superhero known as 'Abomination' according to the database?  Q2:: What superpowers does Abomination have?

LLM only question rating: good.

Fused generated questions: Q1:: What are the names of the superpowers associated with the superhero known as 'Abomination' according to the database?  Q2:: What superpowers does 'Abomination' have?

Fused question rating: good.

---------------------------------------------------

Question 801:

SQL:

SELECT CAST(COUNT(CASE WHEN T2.gender = 'Male' THEN T1.id ELSE NULL END) AS REAL) / COUNT(CASE WHEN T2.gender = 'Female' THEN T1.id ELSE NULL END) FROM superhero AS T1 INNER JOIN gender AS T2 ON T1.gender_id = T2.id

Supplied question: Find the ratio between male superheroes and female superheroes.

Base generated questions: Step 1: The SQL code is performing the following operations: 1. It selects data from two tables - `superhero` and `gender` - using an INNER JOIN on the `gender_id` field of the `superhero` table and the `id` field of the `gender` table. This join operation combines rows from both tables based on a related column between them. 2. It then counts the number of rows from the `superhero` table (using `T1.id` as a reference) where the corresponding `gender` from `T2` is 'Male'. This is done using a CASE statement within the COUNT function, which only counts rows where the condition is met (gender = 'Male'). 3. Similarly, it counts the number of rows where the `gender` is 'Female'. 4. After obtaining the counts for both 'Male' and 'Female' genders, the SQL code casts the count of 'Male' to REAL to ensure that the division operation results in a floating-point number instead of an integer. 5. Finally, it divides the count of 'Male' superheroes by the count of 'Female' superheroes to obtain a ratio.  Question that can be answered by the given SQL code: What is the ratio of the number of male superheroes to the number of female superheroes in the database? Step 2: Rewritten Question: What is the male to female superhero ratio?

Base question rating: good.

Bird only generated questions: Q1:: What is the ratio of the number of male superheroes to the number of female superheroes in the database, based on the gender information associated with each superhero in the 'superhero' and 'gender' tables? Q2:: What is the male to female superhero ratio?

Bird only question rating: good.

LLM only generated questions: Q1:: What is the ratio of the number of male superheroes to the number of female superheroes in the database, according to their respective gender identification in the gender table? Q2:: What is the male to female superhero ratio?

LLM only question rating: good.

Fused generated questions: Q1:: What is the ratio of the number of male superheroes to the number of female superheroes in the database? Q2:: What is the male to female superhero ratio?

Fused question rating: good.

---------------------------------------------------

Question 829:

SQL:

SELECT SUM(CASE WHEN T2.publisher_name = 'DC Comics' THEN 1 ELSE 0 END) - SUM(CASE WHEN T2.publisher_name = 'Marvel Comics' THEN 1 ELSE 0 END) FROM superhero AS T1 INNER JOIN publisher AS T2 ON T1.publisher_id = T2.id

Supplied question: Which publisher created more superheroes: DC or Marvel Comics? Find the difference in the number of superheroes.

Base generated questions: Step 1: The given SQL code is performing the following operations: 1. It selects records from two tables, `superhero` (aliased as T1) and `publisher` (aliased as T2), where the `superhero` table has a `publisher_id` that references the `id` in the `publisher` table. 2. It performs an INNER JOIN on these two tables using the `publisher_id` from the `superhero` table and the `id` from the `publisher` table to match the superheroes with their respective publishers. 3. For each matched record, it checks the `publisher_name` from the `publisher` table: - If the `publisher_name` is 'DC Comics', it counts this as 1. - If the `publisher_name` is 'Marvel Comics', it counts this as -1 (because it's subtracted). - All other `publisher_name` values do not affect the count. 4. It then sums these counts to get two totals: one for 'DC Comics' and one for 'Marvel Comics'. 5. Finally, it subtracts the total count for 'Marvel Comics' from the total count for 'DC Comics'. The question that can be answered by the given SQL code is: "What is the difference in the number of superheroes between DC Comics and Marvel Comics?" Step 2: "How many more superheroes does DC Comics have compared to Marvel Comics?"

Base question rating: good+. Reason: question asks for identity, but only a diffrence in counts is computed.

Bird only generated questions: Q1:: How many more superheroes are published by DC Comics compared to Marvel Comics in the database? Q2:: What is the difference in the number of superheroes between DC Comics and Marvel Comics?

Bird only question rating: good+. Reason: question asks for identity, but only a diffrence in counts is computed.

LLM only generated questions: Q1:: How many more superheroes are published by DC Comics than by Marvel Comics in the database? Q2:: What is the difference in the number of superheroes between DC Comics and Marvel Comics?

LLM only question rating: good+. Reason: question asks for identity, but only a diffrence in counts is computed.

Fused generated questions: Q1:: How many more superheroes are associated with DC Comics than with Marvel Comics in the database? Q2:: What is the difference in the number of superheroes between DC Comics and Marvel Comics?

Fused question rating: good+. Reason: question asks for identity, but only a diffrence in counts is computed.

---------------------------------------------------

Question 544:

SQL:

SELECT T2.DisplayName FROM posts AS T1 INNER JOIN users AS T2 ON T1.LastEditorUserId = T2.Id WHERE T1.Title = 'Examples for teaching: Correlation does not mean causation'

Supplied question: What is the display name of the user who last edited the post "Examples for teaching: Correlation does not mean causation"?

Base generated questions: Q1:: Who is the user that last edited the post with the title 'Examples for teaching: Correlation does not mean causation'? Q2:: Who last edited the 'Correlation vs Causation' teaching post?

Base question rating: good.

Bird only generated questions: Q1:: Who is the display name of the last editor who edited the post with the title 'Examples for teaching: Correlation does not mean causation'? Q2:: What is the name of the user who last edited the 'Correlation does not mean causation' teaching post?

Bird only question rating: good.

LLM only generated questions: Q1:: Who is the display name of the user who last edited the post titled "Examples for teaching: Correlation does not mean causation"? Q2:: Who last edited the "Correlation does not mean causation" teaching post?

LLM only question rating: good.

Fused generated questions: Q1:: Who is the display name of the user who last edited the post titled 'Examples for teaching: Correlation does not mean causation'? Q2:: Who last edited 'Correlation does not mean causation'?

Fused question rating: good.

---------------------------------------------------

Question 565:

SQL:

SELECT IIF(T2.ClosedDate IS NULL, 'NOT well-finished', 'well-finished') AS resylt FROM comments AS T1 INNER JOIN posts AS T2 ON T1.PostId = T2.Id WHERE T1.UserId = 23853 AND T1.CreationDate = '2013-07-12 09:08:18.0'

Supplied question: User No.23853 gave a comment to a post at 9:08:18 on 2013/7/12, was that post well-finished?

Base generated questions: Q1:: Was the post associated with the comment made by user ID 23853 on July 12, 2013, at 09:08:18 finished or not at the time of the query? Q2:: Is the post from July 12, 2013, by user 23853 closed?

Base question rating: good.

Bird only generated questions: Q1:: For the user with ID 23853, what is the status (well-finished or not well-finished) of the post associated with the comment they made on July 12, 2013, at 09:08:18.0, based on whether the post has a closed date? Q2:: What is the completion status of the post commented on by user 23853 on July 12, 2013?

Bird only question rating: good.

LLM only generated questions: Q1:: What is the status of completion ('well-finished' or 'NOT well-finished') for the posts associated with comments made by user ID 23853 on the specific date and time of July 12, 2013, at 09:08:18.0? Q2:: What is the completion status of posts commented on by user 23853 on July 12, 2013, at 9:08 AM?

LLM only question rating: good.

Fused generated questions: Q1:: Was the post associated with the comment made by user 23853 on July 12, 2013, at 09:08:18 well-finished or not? Q2:: Is the post linked to the user 23853's comment from July 12, 2013, closed?

Fused question rating: good.

---------------------------------------------------

Question 576:

SQL:

SELECT T1.DisplayName FROM users AS T1 INNER JOIN comments AS T2 ON T1.Id = T2.UserId WHERE T2.Text = 'thank you user93!'

Supplied question: Name the user that commented 'thank you user93!'

Base generated questions: Q1:: Which users have posted comments with the exact text 'thank you user93!'? Q2:: Who commented 'thank you user93!'?

Base question rating: good.

Bird only generated questions: Q1:: Which users have posted comments containing the exact text 'thank you user93!'? Q2:: Who commented 'thank you user93!'?

Bird only question rating: good.

LLM only generated questions: **Q1::** What is the display name of the user who made a comment with the text 'thank you user93!'? **Q2::** Who commented 'thank you user93!'?

LLM only question rating: good.

Fused generated questions: Q1:: Which users have posted a comment with the exact text 'thank you user93!'? Q2:: Who said 'thank you user93!'?

Fused question rating: good.

---------------------------------------------------

Question 592:

SQL:

SELECT COUNT(UserId) FROM ( SELECT UserId, COUNT(Name) AS num FROM badges GROUP BY UserId ) T WHERE T.num > 5

Supplied question: How many users are awarded with more than 5 badges?

Base generated questions: Q1:: How many users have been awarded more than 5 different types of badges? Q2:: How many users earned over 5 badge types?

Base question rating: good.

Bird only generated questions: Q1:: How many users have earned more than 5 different types of badges? Q2:: How many users received over 5 badges?

Bird only question rating: good.

LLM only generated questions: Q1:: How many users have been awarded more than 5 different badges? Q2:: How many users earned over 5 badges?

LLM only question rating: good.

Fused generated questions: Q1:: How many users have obtained more than five different badges? Q2:: How many users earned over five badges?

Fused question rating: good.

---------------------------------------------------

Question 634:

SQL:

SELECT T1.DisplayName FROM users AS T1 INNER JOIN postHistory AS T2 ON T1.Id = T2.UserId INNER JOIN posts AS T3 ON T2.PostId = T3.Id WHERE T1.DisplayName = 'Harvey Motulsky' OR T1.DisplayName = 'Noah Snyder' GROUP BY T1.DisplayName ORDER BY SUM(T3.ViewCount) DESC LIMIT 1

Supplied question: Among posts by Harvey Motulsky and Noah Snyder, which one has higher popularity?

Base generated questions: Q1:: Which user, either Harvey Motulsky or Noah Snyder, has the highest cumulative view count across all their posts? Q2:: Who has more post views, Harvey Motulsky or Noah Snyder?

Base question rating: good.

Bird only generated questions: Q1:: Which user, either Harvey Motulsky or Noah Snyder, has the highest total view count for all their posts combined in the database? Q2:: Who has more post views, Harvey Motulsky or Noah Snyder?

Bird only question rating: good.

LLM only generated questions: Q1:: Which user, Harvey Motulsky or Noah Snyder, has a display name associated with posts that have the highest total view count? Q2:: Who has more post views, Harvey Motulsky or Noah Snyder?

LLM only question rating: good.

Fused generated questions: Q1:: Which user, either Harvey Motulsky or Noah Snyder, has the highest total view count for all their posts? Q2:: Who has more popular posts, Harvey Motulsky or Noah Snyder?

Fused question rating: good.

---------------------------------------------------

Question 671:

SQL:

SELECT T1.DisplayName FROM users AS T1 INNER JOIN badges AS T2 ON T1.Id = T2.UserId WHERE T2.`Name` = 'Autobiographer' ORDER BY T2.Date LIMIT 1

Supplied question: What is the display name of the user who acquired the first Autobiographer badge?

Base generated questions: Q1:: Which user was the first to receive the 'Autobiographer' badge according to the records in the database? Q2:: Who first earned the 'Autobiographer' badge?

Base question rating: good.

Bird only generated questions: Q1:: Which user was the first to receive the 'Autobiographer' badge according to the recorded data in the system? Q2:: Who first earned the 'Autobiographer' badge?

Bird only question rating: good.

LLM only generated questions: Q1:: Which user was the first to receive the 'Autobiographer' badge according to the recorded dates in the database? Q2:: Who first earned the 'Autobiographer' badge?

LLM only question rating: good.

Fused generated questions: Q1:: Who was the first user to receive the 'Autobiographer' badge according to the recorded dates in the database? Q2:: Who first earned the 'Autobiographer' badge?

Fused question rating: good.

---------------------------------------------------

Question 687:

SQL:

SELECT COUNT(T2.Id) FROM posts AS T1 INNER JOIN comments AS T2 ON T1.Id = T2.PostId GROUP BY T1.Id ORDER BY T1.Score DESC LIMIT 1

Supplied question: How many comments were added to the post with the highest score?

Base generated questions: Q1:: What is the highest number of comments associated with a single post from the 'posts' table, considering only the post with the highest score? Q2:: What's the maximum comment count for the top-scoring post?

Base question rating: good.

Bird only generated questions: Step 1: The given SQL code is performing the following operations: 1. It is joining two tables: 'posts' (aliased as T1) and 'comments' (aliased as T2), on the condition that the 'Id' field in the 'posts' table matches the 'PostId' field in the 'comments' table. 2. It counts the number of comments (T2.Id) associated with each post. 3. It groups the results by the post ID (T1.Id), meaning it calculates the count of comments for each individual post. 4. It orders the grouped results by the 'Score' field of the posts table (T1) in descending order, meaning the post with the highest score will be first. 5. It limits the output to only the first row of the sorted result set. Question that can be answered by the given SQL code: "What is the number of comments on the post with the highest score?" Step 2: Rewritten question in natural language and in fewer words: "How many comments does the top-scoring post have?"

Bird only question rating: good.

LLM only generated questions: Q1:: What is the highest number of comments associated with a single post in the database, considering only the post with the highest score? Q2:: How many comments does the top-scoring post have?

LLM only question rating: good.

Fused generated questions: Q1:: What is the number of comments for the post with the highest score? Q2:: How many comments does the top scored post have?

Fused question rating: good.

---------------------------------------------------

Question 710:

SQL:

SELECT COUNT(T1.id) FROM comments AS T1 INNER JOIN posts AS T2 ON T1.PostId = T2.Id WHERE T2.CommentCount = 1 AND T2.Score = 0

Supplied question: In posts with 1 comment, how many of the comments have 0 score?

Base generated questions: Q1:: How many comments are there on posts that have exactly one comment and a score of zero? Q2:: What is the count of comments on single-comment, zero-score posts?

Base question rating: good.

Bird only generated questions: Q1:: How many comments are there in the database where the associated post has exactly one comment and a score of zero? Q2:: How many comments exist on single-comment, zero-score posts?

Bird only question rating: good.

LLM only generated questions: Q1:: How many comments are associated with posts that have exactly one comment and a score of zero? Q2:: How many comments are on posts with a single comment and no score?

LLM only question rating: good.

Fused generated questions: Q1:: How many comments are there where the associated post has exactly one comment and a score of zero? Q2:: How many comments are linked to posts with only one comment and no upvotes?

Fused question rating: good. Reason: 'no upvotes' is more in line with field metadata.

---------------------------------------------------

Question 346:

SQL:

SELECT T1.id, T1.artist FROM cards AS T1 INNER JOIN legalities AS T2 ON T1.uuid = T2.uuid WHERE T2.status = 'Legal' AND T2.format = 'commander' AND (T1.power IS NULL OR T1.power = '*')

Supplied question: List all the card id and artist with unknown power which are legal for commander play format.

Base generated questions: Q1:: Which cards have a legal status in the Commander format and either do not have a defined power or have a power denoted by an asterisk (*), and what are the names of the artists associated with these cards? Q2:: What are the artist names for Commander-legal cards with undefined or variable power?

Base question rating: good-. Reason: not picking up on * meaning unknown.

Bird only generated questions: Q1:: Which cards, identified by their ID and artist, are legal in the Commander format and either have an undefined power or a power denoted by an asterisk (*)? Q2:: What are the IDs and artists of Commander-legal cards with no specific power?

Bird only question rating: good.

LLM only generated questions: Q1:: Which cards, identified by their ID and artist, have a status of 'Legal' in the 'commander' format and either have a null power value or a power value of '*'? Q2:: What are the IDs and artists of 'commander' legal cards with no power or a power of '*'?

LLM only question rating: good-. Reason: not picking up on * meaning unknown.

Fused generated questions: Q1:: Which cards, identified by their unique ID number, have an artist associated with them and are legal in the Commander format, but either do not have a defined power level or have a power level marked as asterisk (*)? Q2:: What are the IDs and artists of cards with no specific power that are legal in Commander?

Fused question rating: good. Reason: 'no upvotes' is more in line with field metadata.

---------------------------------------------------

Question 366:

SQL:

SELECT T2.format FROM cards AS T1 INNER JOIN legalities AS T2 ON T1.uuid = T2.uuid WHERE T1.name = 'Benalish Knight'

Supplied question: What is the rule of playing card "Benalish Knight"?

Base generated questions: Q1:: What are the formats in which the card named 'Benalish Knight' is legal according to the 'legalities' table? Q2:: In which formats is 'Benalish Knight' a legal card?

Base question rating: good-. Reason: not picking up on format meaning rule, but its in the metadata.

Bird only generated questions: Q1:: Which formats are legal for the card named 'Benalish Knight' according to the database? Q2:: In what formats can 'Benalish Knight' be played?

Bird only question rating: good-. Reason: not picking up on format meaning rule, but its in the metadata.

LLM only generated questions: Q1:: What are the formats in which the card named 'Benalish Knight' is legal to play according to the database, based on the information in the 'legalities' table linked to the 'cards' table through the 'uuid'? Q2:: In which game formats is 'Benalish Knight' a legal card?

LLM only question rating: good-. Reason: not picking up on format meaning rule, but its in the metadata.

Fused generated questions: Q1:: What are the formats of play in which the card named 'Benalish Knight' is legal according to the legalities table? Q2:: In which formats can you play 'Benalish Knight'?

Fused question rating: good-. Reason: not picking up on format meaning rule, but its in the metadata.

---------------------------------------------------

Question 391:

SQL:

SELECT DISTINCT T1.name FROM cards AS T1 INNER JOIN foreign_data AS T2 ON T1.uuid = T2.uuid WHERE T1.originalType = 'Artifact' AND T1.colors = 'B'

Supplied question: Among the Artifact cards, which are black color and comes with foreign languague translation?

Base generated questions: Q1:: What are the names of the distinct cards that are classified as 'Artifact' in their original type and have the color 'B', which also have corresponding entries in the foreign_data table? Q2:: What are the unique names of black artifact cards listed in both the cards and foreign_data tables?

Base question rating: good. Reason: not picking up on foreign_data mening foreign language, but there is no implication that there is a translation.

Bird only generated questions: Q1:: What are the names of distinct cards that are classified as 'Artifact' in their original type and have the color 'B' (Black), which also have corresponding entries in the foreign_data table? Q2:: List the unique names of black Artifact cards with foreign data.

Bird only question rating: good. Reason: not picking up on foreign_data mening foreign language, but there is no implication that there is a translation.

LLM only generated questions: Q1:: What are the names of distinct cards that are classified as 'Artifact' in their original type and have 'B' as their color, according to the data in the 'cards' and 'foreign_data' tables where the 'uuid' fields match? Q2:: What are the unique names of black artifact cards?

LLM only question rating: good. Reason: not picking up on foreign_data mening foreign language, but there is no implication that there is a translation.

Fused generated questions: Q1:: What are the names of all distinct 'Artifact' cards with black color ('B') that have corresponding foreign data entries? Q2:: List the unique black artifact cards with foreign entries.

Fused question rating: good. Reason: not picking up on foreign_data mening foreign language, but there is no implication that there is a translation.

---------------------------------------------------

Question 409:

SQL:

SELECT COUNT(T1.id) FROM cards AS T1 INNER JOIN legalities AS T2 ON T1.uuid = T2.uuid INNER JOIN rulings AS T3 ON T1.uuid = T3.uuid WHERE T2.format = 'premodern' AND T3.text = 'This is a triggered mana ability.' AND T1.Side IS NULL

Supplied question: Indicates the number of cards with pre-modern format, ruling text "This is a triggered mana ability." that do not have multiple faces.

Base generated questions: Q1:: How many cards in the 'premodern' format do not have a specified side and have a ruling text that states 'This is a triggered mana ability.'? Q2:: How many 'premodern' cards with no side have a ruling about being a triggered mana ability?

Base question rating: good+. Reason: poor grammer in the supplied question.

Bird only generated questions: Q1:: How many cards that have no side specified are there in the 'premodern' format with rulings that state 'This is a triggered mana ability'? Q2:: How many 'premodern' cards without a side have a ruling mentioning a triggered mana ability?

Bird only question rating: good+. Reason: poor grammer in the supplied question.

LLM only generated questions: **Q1::** How many cards from the 'cards' table, that are involved in the 'premodern' format as per the 'legalities' table and have a ruling stating 'This is a triggered mana ability' according to the 'rulings' table, do not have a specified side? **Q2::** How many premodern cards with no specified side have a ruling that describes a triggered mana ability?

LLM only question rating: good+. Reason: poor grammer in the supplied question.

Fused generated questions: Q1:: How many cards that do not have multiple faces are associated with the 'premodern' format and have a ruling text that states 'This is a triggered mana ability'? Q2:: How many single-faced 'premodern' cards have a ruling about a triggered mana ability?

Fused question rating: good+. Reason: poor grammer in the supplied question.

---------------------------------------------------

Question 424:

SQL:

SELECT CAST(SUM(CASE WHEN isTextless = 1 AND layout = 'normal' THEN 1 ELSE 0 END) AS REAL) * 100 / COUNT(*) FROM cards

Supplied question: What proportion of cards do not have a text box with a normal layout?

Base generated questions: Q1:: What is the percentage of 'normal' layout cards without text out of the total number of cards in the 'cards' table? Q2:: What percent of all cards have a normal layout and no text?

Base question rating: good.

Bird only generated questions: Q1:: What is the percentage of 'cards' that have the 'isTextless' attribute set to 1 and the 'layout' attribute set to 'normal' out of all the cards in the database? Q2:: What percentage of all cards are textless with a normal layout?

Bird only question rating: good.

LLM only generated questions: Q1:: What is the percentage of cards that are both textless (isTextless = 1) and have a 'normal' layout out of the total number of cards in the database? Q2:: What percent of the cards are textless with a normal layout?

LLM only question rating: good.

Fused generated questions: Q1:: What is the percentage of cards in the 'cards' table that are textless (isTextless = 1) and have a 'normal' layout? Q2:: What percent of normal layout cards are textless?

Fused question rating: good.

---------------------------------------------------

Question 466:

SQL:

SELECT COUNT(T1.id) FROM cards AS T1 INNER JOIN set_translations AS T2 ON T2.setCode = T1.setCode WHERE T2.translation = 'Hauptset Zehnte Edition' AND T1.artist = 'Adam Rex'

Supplied question: Among the cards in the set "Hauptset Zehnte Edition", how many of them are designed by Adam Rex?

Base generated questions: Q1:: How many cards illustrated by Adam Rex are there in the set with the translation name 'Hauptset Zehnte Edition'? Q2:: How many 'Hauptset Zehnte Edition' cards did Adam Rex illustrate?

Base question rating: good. Reason: card_games.cards.artist referes to the illustrator, not the designer.

Bird only generated questions: Q1:: How many cards illustrated by Adam Rex are part of the set whose translation is 'Hauptset Zehnte Edition'? Q2:: How many 'Hauptset Zehnte Edition' cards did Adam Rex illustrate?

Bird only question rating: good. Reason: card_games.cards.artist referes to the illustrator, not the designer.

LLM only generated questions: Q1:: How many cards illustrated by Adam Rex are there in the set with the translation 'Hauptset Zehnte Edition'? Q2:: How many 'Hauptset Zehnte Edition' cards did Adam Rex illustrate?

LLM only question rating: good. Reason: card_games.cards.artist referes to the illustrator, not the designer.

Fused generated questions: Q1:: How many cards illustrated by Adam Rex are there in the database that belong to the set with the translation 'Hauptset Zehnte Edition'? Q2:: How many 'Hauptset Zehnte Edition' cards were illustrated by Adam Rex?

Fused question rating: good. Reason: card_games.cards.artist referes to the illustrator, not the designer.

---------------------------------------------------

Question 477:

SQL:

SELECT T1.artist FROM cards AS T1 INNER JOIN sets AS T2 ON T2.code = T1.setCode WHERE (T2.name = 'Coldsnap' AND T1.artist = 'Chippy') OR (T2.name = 'Coldsnap' AND T1.artist = 'Aaron Miller') OR (T2.name = 'Coldsnap' AND T1.artist = 'Jeremy Jarvis') GROUP BY T1.artist

Supplied question: Which of these artists have designed a card in the set Coldsnap, Jeremy Jarvis, Aaron Miller or Chippy?

Base generated questions: Q1:: Which artists named 'Chippy', 'Aaron Miller', or 'Jeremy Jarvis' have created art for cards in the 'Coldsnap' set? Q2:: Who are the 'Coldsnap' card artists among Chippy, Aaron Miller, and Jeremy Jarvis?

Base question rating: good.

Bird only generated questions: Q1:: Which artists have contributed to the 'Coldsnap' set whose names are either 'Chippy', 'Aaron Miller', or 'Jeremy Jarvis'? Q2:: Who are the 'Coldsnap' set artists named 'Chippy', 'Aaron Miller', or 'Jeremy Jarvis'?

Bird only question rating: good.

LLM only generated questions: **Q1::** Which artists named 'Chippy', 'Aaron Miller', or 'Jeremy Jarvis' have contributed to the 'Coldsnap' set in the database? **Q2::** Which artists from 'Chippy', 'Aaron Miller', and 'Jeremy Jarvis' have artwork in the 'Coldsnap' set?

LLM only question rating: good.

Fused generated questions: Q1:: Which artists named 'Chippy', 'Aaron Miller', or 'Jeremy Jarvis' have illustrated cards in the 'Coldsnap' set? Q2:: Who are the artists from 'Coldsnap' set that are either 'Chippy', 'Aaron Miller', or 'Jeremy Jarvis'?

Fused question rating: good.

---------------------------------------------------

Question 487:

SQL:

SELECT CAST(SUM(CASE WHEN T1.cardKingdomFoilId IS NOT NULL AND T1.cardKingdomId IS NOT NULL THEN 1 ELSE 0 END) AS REAL) * 100 / COUNT(T1.id) FROM cards AS T1 INNER JOIN sets AS T2 ON T2.code = T1.setCode WHERE T2.name = 'Coldsnap'

Supplied question: What is the percentage of incredibly powerful cards in the set Coldsnap?

Base generated questions: Q1:: What is the percentage of cards from the 'Coldsnap' set that have both a 'cardKingdomFoilId' and a 'cardKingdomId' associated with them in the database? Q2:: What percentage of 'Coldsnap' set cards have both foil and non-foil versions available on Card Kingdom?

Base question rating: good+. Reason: supplied question does not match the SQL.

Bird only generated questions: Q1:: What is the percentage of cards in the 'Coldsnap' set that have both a Card Kingdom foil ID and a Card Kingdom ID? Q2:: What percentage of 'Coldsnap' cards have both foil and non-foil versions available on Card Kingdom?

Bird only question rating: good+. Reason: supplied question does not match the SQL.

LLM only generated questions: Q1:: What is the percentage of cards in the 'Coldsnap' set that have both a 'cardKingdomFoilId' and a 'cardKingdomId'? Q2:: What percentage of 'Coldsnap' cards have both foil and non-foil versions on Card Kingdom?

LLM only question rating: good+. Reason: supplied question does not match the SQL.

Fused generated questions: Q1:: What percentage of cards from the 'Coldsnap' set have both a unique identifier for the standard version and a unique identifier for the foil version as assigned by Card Kingdom? Q2:: What is the percentage of 'Coldsnap' cards with identifiers for both standard and foil versions from Card Kingdom?

Fused question rating: good+. Reason: supplied question does not match the SQL.

---------------------------------------------------

Question 195:

SQL:

SELECT T.bond_type FROM ( SELECT bond_type, COUNT(bond_id) FROM bond GROUP BY bond_type ORDER BY COUNT(bond_id) DESC LIMIT 1 ) AS T

Supplied question: What is the most common bond type?

Base generated questions: Step 1: The given SQL code is performing the following steps: 1. It is selecting two columns from the 'bond' table: 'bond_type' and the count of 'bond_id'. 2. It groups the results by 'bond_type' so that it can calculate the count of 'bond_id' for each 'bond_type'. 3. It orders the grouped results in descending order by the count of 'bond_id' to find out which 'bond_type' has the highest number of bonds. 4. It limits the results to only the top row, which effectively means it selects the 'bond_type' with the highest count of bonds. 5. It then creates a subquery (alias as T) and from this subquery, it selects the 'bond_type'. Question that can be answered by the given SQL code: What is the type of bond that has the highest number of entries in the bond table? Step 2: What is the most common bond type?

Base question rating: good.

Bird only generated questions: Q1:: What is the most common type of bond in the bond table? Q2:: Which bond type occurs most frequently?

Bird only question rating: good.

LLM only generated questions: Q1:: What is the most common type of bond in the bond table, based on the highest occurrence of bond IDs? Q2:: What is the most frequent bond type?

LLM only question rating: good.

Fused generated questions: Q1:: What is the most common type of chemical bond present in the molecules recorded in the bond table? Q2:: Which bond type occurs most frequently?

Fused question rating: good.

---------------------------------------------------

Question 207:

SQL:

SELECT DISTINCT T1.element FROM atom AS T1 INNER JOIN bond AS T2 ON T1.molecule_id = T2.molecule_id INNER JOIN connected AS T3 ON T1.atom_id = T3.atom_id WHERE T2.bond_type = '='

Supplied question: What elements are in a double type bond?

Base generated questions: Q1:: What are the distinct elements that are part of molecules where at least one bond is a double bond (represented by '=')? Q2:: Which unique elements are in molecules with double bonds?

Base question rating: good.

Bird only generated questions: Q1:: What are the distinct elements in the 'atom' table that are part of molecules where at least one bond is a double bond ('=') and these atoms are also present in the 'connected' table? Q2:: Which elements form double bonds in connected molecules?

Bird only question rating: good.

LLM only generated questions: Q1:: What are the distinct elements from the 'atom' table that are part of molecules where there is at least one double bond ('=') present, according to the 'bond' table, and have a corresponding entry in the 'connected' table? Q2:: Which elements are in molecules with double bonds?

LLM only question rating: good.

Fused generated questions: Q1:: Which chemical elements are part of molecules that have at least one double bond? Q2:: What elements are in molecules with double bonds?

Fused question rating: good.

---------------------------------------------------

Question 219:

SQL:

SELECT CAST(COUNT(DISTINCT CASE WHEN T2.label = '+' THEN T2.molecule_id ELSE NULL END) AS REAL) * 100 / COUNT(DISTINCT T2.molecule_id) FROM atom AS T1 INNER JOIN molecule AS T2 ON T1.molecule_id = T2.molecule_id INNER JOIN bond AS T3 ON T2.molecule_id = T3.molecule_id WHERE T3.bond_type = '#'

Supplied question: What is the percentage of carcinogenic molecules in triple type bonds?

Base generated questions: Q1:: What is the percentage of distinct molecules labeled with a '+' that have at least one triple bond, out of the total number of distinct molecules that have at least one triple bond? Q2:: What percentage of molecules with a triple bond are positively labeled?

Base question rating: good-. Reason: not picking up on meaning of label.

Bird only generated questions: Q1:: What is the percentage of distinct molecules with a positive label ('+') among all distinct molecules that have at least one atom involved in a bond with a bond type of '#'? Q2:: What percentage of molecules with a triple bond are positively labeled?

Bird only question rating: good-. Reason: not picking up on meaning of label.

LLM only generated questions: **Q1::** What is the percentage of molecules that have a label of '+' among all the molecules that contain at least one bond of type '#'? **Q2::** What percentage of molecules with a '#' bond type are labeled with '+'?

LLM only question rating: bad+. Reason: not picking up on the meanings of the literals.

Fused generated questions: Q1:: What is the percentage of unique molecules that contain a triple bond and are carcinogenic relative to the total number of unique molecules that contain a triple bond? Q2:: What percentage of molecules with a triple bond are carcinogenic?

Fused question rating: good.

---------------------------------------------------

Question 231:

SQL:

SELECT T.bond_type FROM ( SELECT T1.bond_type, COUNT(T1.molecule_id) FROM bond AS T1 WHERE T1.molecule_id = 'TR010' GROUP BY T1.bond_type ORDER BY COUNT(T1.molecule_id) DESC LIMIT 1 ) AS T

Supplied question: Which bond type accounted for the majority of the bonds found in molecule TR010 and state whether or not this molecule is carcinogenic?

Base generated questions: Step 1: What is the most common type of bond found in the molecule with the ID 'TR010' in the bond table? Step 2: What is the most frequent bond type in molecule 'TR010'?

Base question rating: good+. Reason: supplied question does not match the SQL.

Bird only generated questions: Step 1: What is the most common type of bond in the molecule with the ID 'TR010' according to the bond records in the database? Step 2: What is the most frequent bond type in molecule 'TR010'?

Bird only question rating: good+. Reason: supplied question does not match the SQL.

LLM only generated questions: Q1:: What is the most common type of bond in molecule 'TR010' according to the records in the bond table? Q2:: What is the most frequent bond type for molecule 'TR010'?

LLM only question rating: good+. Reason: supplied question does not match the SQL.

Fused generated questions: Q1:: What is the most common type of chemical bond in the molecule with the identifier 'TR010'? Q2:: What is the predominant bond type in molecule 'TR010'?

Fused question rating: good+. Reason: supplied question does not match the SQL.

---------------------------------------------------

Question 242:

SQL:

SELECT DISTINCT T2.molecule_id FROM atom AS T1 INNER JOIN molecule AS T2 ON T1.molecule_id = T2.molecule_id WHERE SUBSTR(T1.atom_id, -2) BETWEEN '21' AND '25' AND T2.label = '+'

Supplied question: Among all the atoms from 21 to 25, list all the molecules that are carcinogenic.

Base generated questions: Q1:: Which molecule IDs are associated with atoms whose IDs end in a number between 21 and 25, and also have a molecule label of '+'? Q2:: What are the IDs of molecules with a '+' label and containing atoms ending in 21 to 25?

Base question rating: good-. Reason: not picking up on meaning of label.

Bird only generated questions: Q1:: Which molecule IDs are associated with atoms whose IDs end in a number between 21 and 25 and are also part of molecules labeled with a positive charge? Q2:: What are the IDs of positively charged molecules with atom IDs ending in 21 to 25?

Bird only question rating: good-. Reason: not picking up on meaning of label.

LLM only generated questions: Q1:: Which molecule IDs are associated with atoms whose IDs end in a number between 21 and 25, and also have a molecule label of '+'? Q2:: What are the IDs of molecules with '+' label and atoms ending in 21 to 25?

LLM only question rating: good-. Reason: not picking up on meaning of label.

Fused generated questions: Q1:: Which unique identifiers (molecule_id) belong to molecules that are carcinogenic and have at least one atom whose identifier ends with a number between 21 and 25? Q2:: What are the IDs of carcinogenic molecules with atoms numbered 21 to 25?

Fused question rating: good.

---------------------------------------------------

Question 249:

SQL:

SELECT T2.element FROM connected AS T1 INNER JOIN atom AS T2 ON T1.atom_id = T2.atom_id WHERE T1.bond_id = 'TR144_8_19'

Supplied question: What are the elements of the atoms of TR144_8_19?

Base generated questions: Q1:: What are the names of the elements connected by the bond with the ID 'TR144_8_19' in the database? Q2:: Which elements are bonded by 'TR144_8_19'?

Base question rating: good. Reason: supplied question is vague.

Bird only generated questions: Q1:: What are the elements associated with the atom entries in the 'atom' table that are connected through the bond with ID 'TR144_8_19' in the 'connected' table? Q2:: Which elements are bonded by 'TR144_8_19'?

Bird only question rating: good. Reason: supplied question is vague.

LLM only generated questions: Q1:: What are the elements associated with the atom_ids that are connected through the bond_id 'TR144_8_19'? Q2:: Which elements are connected by bond 'TR144_8_19'?

LLM only question rating: good. Reason: supplied question is vague.

Fused generated questions: Q1:: What are the chemical element symbols of the atoms that are connected by the bond with the identifier 'TR144_8_19'? Q2:: Which elements are bonded in 'TR144_8_19'?

Fused question rating: good. Reason: supplied question is vague.

---------------------------------------------------

Question 273:

SQL:

SELECT CAST(COUNT( CASE WHEN T1.element = 'cl' THEN T1.element ELSE NULL END) AS REAL) * 100 / COUNT(T1.element) FROM atom AS T1 INNER JOIN molecule AS T2 ON T1.molecule_id = T2.molecule_id WHERE T2.label = '+'

Supplied question: What is the percentage of element chlorine in carcinogenic molecules?

Base generated questions: Q1:: What is the percentage of 'cl' elements within all the elements associated with molecules that have a label of '+'? Q2:: What percentage of elements are chlorine in positively labeled molecules?

Base question rating: good-. Reason: not picking up on meaning of label.

Bird only generated questions: Q1:: What is the percentage of 'cl' elements within the atoms associated with molecules that have a label of '+'? Q2:: What percentage of atoms in positively labeled molecules are chlorine?

Bird only question rating: good-. Reason: not picking up on meaning of label.

LLM only generated questions: **Q1::** What is the percentage of atoms labeled as 'cl' in molecules with a label of '+'? **Q2::** What percentage of atoms are chlorine in positively labeled molecules?

LLM only question rating: good-. Reason: not picking up on meaning of label.

Fused generated questions: Q1:: What is the percentage of chlorine atoms among all atoms in carcinogenic molecules? Q2:: What percent of atoms are chlorine in carcinogenic compounds?

Fused question rating: good.

---------------------------------------------------

Question 12:

SQL:

```
SELECT MAX(CAST(T1.`Free Meal Count (Ages 5-17)` AS REAL) / T1.`Enrollment (Ages 5-17)`) FROM frpm AS T1 INNER JOIN satscores AS T2 ON T1.CDSCode = T2.cds WHERE CAST(T2.NumGE1500 AS REAL) / T2.NumTstTakr > 0.3
```

Supplied question: Among the schools with an SAT excellence rate of over 0.3, what is the highest eligible free rate for students aged 5-17?

Base generated questions: Q1:: What is the highest ratio of free meals provided to students aged 5-17 to the total enrollment of students aged 5-17 in schools where more than 30% of students taking the SAT scored 1500 or higher? Q2:: What is the maximum percentage of students aged 5-17 receiving free meals in schools with over 30% of SAT scores above 1500?

Base question rating: good-. Reason: Not picking up on sat>=1500 meaning excellent.

Bird only generated questions: Q1:: What is the highest ratio of free meals provided to students aged 5-17 to the total enrollment of students aged 5-17 in schools where more than 30% of students taking the SAT scored 1500 or higher? Q2:: What is the maximum percentage of students aged 5-17 receiving free meals in schools with over 30% of SAT scores above 1500?

Bird only question rating: good-. Reason: Not picking up on sat>=1500 meaning excellent.

LLM only generated questions: Q1:: What is the highest ratio of students aged 5-17 receiving free meals to the total enrollment of students aged 5-17 in schools where more than 30% of test takers scored above 1500 on the SAT? Q2:: What is the maximum proportion of students aged 5-17 on free meals in schools where over 30% of SAT takers score above 1500?

LLM only question rating: good-. Reason: Not picking up on sat>=1500 meaning excellent.

Fused generated questions: Q1:: What is the highest ratio of students eligible for free meals to total enrollment for schools where more than 30% of SAT test takers scored 1500 or higher? Q2:: What is the maximum free meal eligibility rate among schools with over 30% SAT excellence?

Fused question rating: good.

---------------------------------------------------

Question 27:

SQL:

```
SELECT T2.School, T1.AvgScrWrite, T2.Phone FROM schools AS T2 LEFT JOIN satscores AS T1 ON T2.CDSCode = T1.cds WHERE strftime('%Y', T2.OpenDate) > '1991' OR strftime('%Y', T2.ClosedDate) < '2000'
```

Supplied question: What is the average score in writing for the schools that were opened after 1991 or closed before 2000? List the school names along with the score. Also, list the communication number of the schools if there is any.

Base generated questions: Q1:: Which schools that opened after 1991 or closed before 2000 have associated average writing SAT scores and phone numbers according to the schools and satscores tables? Q2:: What are the phone numbers and average SAT writing scores for schools opened post-1991 or closed pre-2000?

Base question rating: good-. Reason: Not asking for school name.

Bird only generated questions: Q1:: Which schools that opened after 1991 or closed before 2000 have available average writing SAT scores and contact phone numbers? Q2:: What are the SAT writing scores and phone numbers for schools opened post-1991 or closed pre-2000?

Bird only question rating: good-. Reason: Not asking for school name.

LLM only generated questions: Q1:: Which schools that opened after 1991 or closed before 2000 have associated average SAT writing scores and contact phone numbers? Q2:: What are the names, average SAT writing scores, and phone numbers of schools opened post-1991 or closed pre-2000?

LLM only question rating: good.

Fused generated questions: Q1:: What are the names, average SAT writing scores, and contact phone numbers of schools that were opened after 1991 or closed before 2000? Q2:: List the schools with their SAT writing scores and phone numbers, which opened post-1991 or closed pre-2000.

Fused question rating: good.

---------------------------------------------------

Question 39:

SQL:

```
SELECT AVG(T1.NumTstTakr) FROM satscores AS T1 INNER JOIN schools AS T2 ON T1.cds = T2.CDSCode WHERE strftime('%Y', T2.OpenDate) = '1980' AND T2.County = 'Fresno'
```

Supplied question: What is the average number of test takers from Fresno schools that opened between 1/1/1980 and 12/31/1980?

Base generated questions: Q1:: What is the average number of test takers for the SAT from schools in Fresno County that were opened in the year 1980? Q2:: What's the average SAT participation for schools opened in 1980 in Fresno County?

Base question rating: good. Reason: supplied question is vague.

Bird only generated questions: Q1:: What is the average number of test takers for the SAT from schools in Fresno County that were opened in the year 1980? Q2:: What is Fresno's 1980 school average SAT test takers?

Bird only question rating: good. Reason: supplied question is vague.

LLM only generated questions: Q1:: What is the average number of test takers for the SAT from schools in Fresno County that were opened in the year 1980? Q2:: What's the average SAT participation for Fresno schools founded in 1980?

LLM only question rating: good. Reason: supplied question is vague.

Fused generated questions: Q1:: What is the average number of students who took the SAT at schools in Fresno County that were opened in the year 1980? Q2:: What's the average SAT participation at Fresno schools started in 1980?

Fused question rating: good. Reason: supplied question is vague.

---------------------------------------------------

Question 48:

SQL:

SELECT CAST(SUM(CASE WHEN DOC = 54 THEN 1 ELSE 0 END) AS REAL) / SUM(CASE WHEN DOC = 52 THEN 1 ELSE 0 END) FROM schools WHERE StatusType = 'Merged' AND County = 'Orange'

Supplied question: What is the ratio of merged Unified School District schools in Orange County to merged Elementary School District schools?

Base generated questions: Q1:: What is the ratio of the number of schools with DOC code 54 to the number of schools with DOC code 52 in Orange County where the StatusType is 'Merged'? Q2:: What is the ratio of type 54 to type 52 merged schools in Orange County?

Base question rating: good-. Reason: not picking up on school district code.

Bird only generated questions: Q1:: What is the ratio of the number of schools with a DOC code of 54 to the number of schools with a DOC code of 52 in Orange County that have a status type of 'Merged'? Q2:: What is the 'Merged' schools ratio of DOC 54 to DOC 52 in Orange County?

Bird only question rating: good-. Reason: not picking up on school district code.

LLM only generated questions: Q1:: What is the ratio of the number of schools with document type 54 to the number of schools with document type 52 in Orange County among those that have a status type of 'Merged'? Q2:: What is the ratio of type 54 to type 52 merged schools in Orange County?

LLM only question rating: good-. Reason: not picking up on school district code.

Fused generated questions: Q1:: What is the ratio of the number of merged unified school districts to the number of merged elementary school districts in Orange County? Q2:: What's the ratio of merged unified to elementary districts in Orange County?

Fused question rating: good.

---------------------------------------------------

Question 82:

SQL:

SELECT GSoffered FROM schools ORDER BY ABS(longitude) DESC LIMIT 1

Supplied question: What is the grade span offered in the school with the highest longitude?

Base generated questions: Q1:: Which school offers a Graduate School program that is located furthest east or west from the Prime Meridian, based on the absolute value of its longitude? Q2:: What is the Graduate School offering of the school located at the extreme longitude?

Base question rating: bad+. Reason: not picking up on the meaning of GSoffered.

Bird only generated questions: Q1:: What is the grade span offered by the school that is located the farthest east or west from the Prime Meridian? Q2:: Which school farthest from the Prime Meridian offers what grade levels?

Bird only question rating: good.

LLM only generated questions: Q1:: Which school offers a graduate program and is located at the point with the greatest absolute value of longitude? Q2:: What is the graduate program at the school furthest east or west?

LLM only question rating: bad+. Reason: not picking up on meaning of GSoffered.

Fused generated questions: Q1:: What is the grade span offered by the school located furthest east or west from the prime meridian? Q2:: Which school has the widest range of grades and is furthest from the prime meridian?

Fused question rating: good. Reason: awkward but accurate.

---------------------------------------------------

Question 93:

SQL:

SELECT COUNT(T1.client_id) FROM client AS T1 INNER JOIN district AS T2 ON T1.district_id = T2.district_id WHERE T1.gender = 'M' AND T2.A3 = 'north Bohemia' AND T2.A11 > 8000

Supplied question: How many male customers who are living in North Bohemia have average salary greater than 8000?

Base generated questions: Q1:: How many male clients are there who live in a district classified as 'north Bohemia' where the district's attribute A11 (which could represent a specific demographic or economic feature) has a value greater than 8000? Q2:: How many men from the North Bohemia district with an A11 value over 8000 are clients?

Base question rating: good-. Reason: not picking up on meaning of A11.

Bird only generated questions: Q1:: How many male clients are there in the 'north Bohemia' region of the 'district' table where the socio-economic indicator 'A11' is greater than 8000? Q2:: How many men from North Bohemia districts with a high socio-economic status are clients?

Bird only question rating: good-. Reason: not picking up on meaning of A11.

LLM only generated questions: Q1:: How many male clients are there in the 'north Bohemia' region where the population of the district is greater than 8000? Q2:: How many men from districts in north Bohemia with populations over 8000 are clients?

LLM only question rating: bad. Reason: misses the meaning.

Fused generated questions: Q1:: How many male clients are associated with branches located in the 'north Bohemia' region where the average salary of the district is greater than 8000? Q2:: How many men are clients in 'north Bohemia' with district salaries over 8000?

Fused question rating: good.

---------------------------------------------------

Question 112:

SQL:

SELECT T1.A2 FROM district AS T1 INNER JOIN client AS T2 ON T1.district_id = T2.district_id WHERE T2.birth_date = '1976-01-29' AND T2.gender = 'F'

Supplied question: For the female client who was born in 1976/1/29, which district did she opened her account?

Base generated questions: Q1:: Which districts do female clients born on January 29, 1976, belong to according to the database? Q2:: What are the districts of female clients born on 29th January 1976?

Base question rating: good+. Reason: supplied question presupposes exactly one client.

Bird only generated questions: Q1:: Which districts do female clients born on January 29, 1976, belong to? Q2:: What are the districts of female clients born on 1/29/1976?

Bird only question rating: good+. Reason: supplied question presupposes exactly one client.

LLM only generated questions: Q1:: What are the names of the districts (A2) where female clients born on January 29, 1976, are located according to the database? Q2:: Which districts have female clients born on 29th January 1976?

LLM only question rating: good+. Reason: supplied question presupposes exactly one client.

Fused generated questions: Q1:: What are the names of the districts where female clients who were born on January 29, 1976, are associated with? Q2:: Which districts have female clients born on 29th January 1976?

Fused question rating: good+. Reason: supplied question presupposes exactly one client.

---------------------------------------------------

Question 128:

SQL:

SELECT T2.A2, COUNT(T1.client_id) FROM client AS T1 INNER JOIN district AS T2 ON T1.district_id = T2.district_id WHERE T1.gender = 'F' GROUP BY T2.district_id, T2.A2 ORDER BY COUNT(T1.client_id) DESC LIMIT 9

Supplied question: List the top nine districts, by descending order, from the highest to the lowest, the number of female account holders.

Base generated questions: Q1:: Which are the top 9 districts with the highest number of female clients, and how many female clients are there in each of these districts? Q2:: What are the top 9 districts by female client count?

Base question rating: good. Reason: supplied question is vague.

Bird only generated questions: Q1:: What are the top 9 districts with the highest number of female clients, along with the count of female clients in each of those districts? Q2:: Which 9 districts have the most female clients?

Bird only question rating: good. Reason: supplied question is vague.

LLM only generated questions: Q1:: Which nine districts have the highest number of female clients, and how many female clients are there in each of these districts? Q2:: What are the top nine districts with the most female clients?

LLM only question rating: good. Reason: supplied question is vague.

Fused generated questions: Q1:: What are the top 9 districts with the highest number of female clients, along with the count of female clients in each of those districts? Q2:: Which 9 districts have the most female clients?

Fused question rating: good. Reason: supplied question is vague.

---------------------------------------------------

Question 138:

SQL:

SELECT COUNT(T1.client_id) FROM client AS T1 INNER JOIN district AS T2 ON T1.district_id = T2.district_id WHERE T1.gender = 'M' AND T2.A15 = (SELECT T3.A15 FROM district AS T3 ORDER BY T3.A15 DESC LIMIT 1, 1)

Supplied question: In the branch where the second-highest number of crimes were committed in 1995 occurred, how many male clients are there?

Base generated questions: Q1:: How many male clients are there in the district with the second-highest value of A15? Q2:: How many men live in the district with the second-highest A15?

Base question rating: good-. Reason: not picking up on meaning of A15.

Bird only generated questions: Q1:: How many male clients are there in the district with the second-highest A15 value? Q2:: How many men live in the district with the second-highest A15?

Bird only question rating: good-. Reason: not picking up on meaning of A15.

LLM only generated questions: Q1:: How many male clients are there in the district with the second highest value of the A15 attribute? Q2:: How many men live in the district with the second highest A15 value?

LLM only question rating: good-. Reason: not picking up on meaning of A15.

Fused generated questions: Q1:: How many male clients are associated with the district that has the second-highest number of committed crimes in 1995? Q2:: How many men are from the district with the second-most crimes in 1995?

Fused question rating: good.

---------------------------------------------------

Question 169:

SQL:

SELECT CAST((SUM(CASE WHEN STRFTIME('%Y', T1.date) = '1997' THEN T1.amount ELSE 0 END) - SUM(CASE WHEN STRFTIME('%Y', T1.date) = '1996' THEN T1.amount ELSE 0 END)) AS REAL) * 100 / SUM(CASE WHEN STRFTIME('%Y', T1.date) = '1996' THEN T1.amount ELSE 0 END) FROM loan AS T1 INNER JOIN account AS T2 ON T1.account_id = T2.account_id INNER JOIN disp AS T3 ON T3.account_id = T2.account_id INNER JOIN client AS T4 ON T4.client_id = T3.client_id WHERE T4.gender = 'M' AND T3.type = 'OWNER'

Supplied question: What was the growth rate of the total amount of loans across all accounts for a male client between 1996 and 1997?

Base generated questions: Q1:: What is the percentage change in the sum of loan amounts from male clients who are owners of their accounts between the years 1996 and 1997? Q2:: What is the loan amount percentage increase or decrease for male account owners from 1996 to 1997?

Base question rating: good. Reason: SQL has stability problem.

Bird only generated questions: Q1:: What is the percentage change in the sum of loan amounts for male clients who are owners of their accounts from the year 1996 to 1997? Q2:: What's the loan amount percentage increase for male account owners from 1996 to 1997?

Bird only question rating: good. Reason: SQL has stability problem.

LLM only generated questions: Q1:: What is the percentage change in the sum of loan amounts from the year 1996 to 1997 for male clients who are owners of their accounts? Q2:: What is the percentage increase or decrease in loans taken by male account owners from 1996 to 1997?

LLM only question rating: good. Reason: SQL has stability problem.

Fused generated questions: Q1:: What is the percentage change in the total amount of loans approved for male clients who are owners of the account between the years 1996 and 1997? Q2:: What is the percentage increase or decrease in approved loan amounts for male account owners from 1996 to 1997?

Fused question rating: good. Reason: SQL has stability problem.

---------------------------------------------------